\documentclass[dvipsnames]{article}

\usepackage[margin=1in]{geometry}

\usepackage{graphicx,amsmath,upgreek,bm}
\usepackage{colortbl}
\usepackage[rgb, table]{xcolor}
\usepackage{booktabs}
\usepackage{makecell}
\usepackage{pifont}
\usepackage{todonotes}
\usepackage[toc,page]{appendix}
\usepackage{multirow}
\usepackage{algorithm}
\usepackage[noend]{algpseudocode}
\usepackage{arydshln}
\makeatletter
  \renewcommand*\env@matrix[1][*\c@MaxMatrixCols c]{%
    \hskip -\arraycolsep
    \let\@ifnextchar\new@ifnextchar
  \array{#1}}
\makeatother
\newcommand{\cmark}{\textcolor{OliveGreen}{\ding{51}}}%
\newcommand{\xmark}{\textcolor{BrickRed}{\ding{55}}}%

\makeatletter
\def\BState{\State\hskip-\ALG@thistlm}
\makeatother
\definecolor{light-gray}{gray}{0.9}

\DeclareMathOperator{\atantwo}{atan2}

\graphicspath{{./figures/}{../figures}}

\title{Exact and Numerically Stable Expressions for Euler-Bernoulli and Timoshenko Beam Modes}

\author{Firas A.~Khasawneh\footnote{khasawn3@egr.msu.edu}\, and Daniel Segalman\footnote{segalman@egr.msu.edu}, \\ Department
  of Mechanical Engineering \\ Michigan State University}

  \date{}

\begin{document}
\maketitle

\begin{abstract}
In this work we present a general procedure for deriving exact, analytical, and numerically stable expressions for the characteristic equations and the eigenmodes of the Timoshenko and the Euler-Bernoulli beam models.
This work generalizes the approach recently described in Gon\c{c}alves et al.\ (P.J.P.~Goncalves, A.~Peplow, M.J.~Brennan, Exact expressions for numerical evaluation of high order modes of vibration in uniform Euler-Bernoulli beams, Applied Acoustics 141 (2018) 371--373), which allows the numerical stabilization for the case of the Timoshenko beam model.
Our results enable the reliable computation of the eigenvalues and the eigenmodes of both beam models for any number of modes.
In addition to presenting the necessary details for stabilizing the solutions to the eigenvalue problem for the two beam models, we also tabulate the results for a large number of the common boundary conditions so that one can compare the predictions of all those models. 
Therefore, another contribution of our work is the presentation of both the conventional as well as the novel, numerically stabilized results for both the Euler-Bernoulli and the Timoshenko beam models in one manuscript with consistent notation, and for the most common boundary conditions. 
The code for the stabilized Timoshenko expressions as well as for the finite element verification are made available through the Mendeley Data repository http://dx.doi.org/10.17632/r275tx2yp8.1.
\end{abstract}

%!TEX root = ../Timoshenko.tex
\pdfoutput=1
\section{Introduction}
The elastic deformation of continuous bodies can generally
be studied using the theory of elasticity.  However, when
certain loading conditions are applied to special structural
elements, it is possible to simplify the analysis by
utilizing the kinematics of the deformation and making some
assumptions on the resulting strains.  One example of these
simplifications is the beam structural element which has
been widely used to model a variety of natural and
engineered components
\cite{Erturk2006,Jiang2010,Claeyssen2013,Martin2018}.
Several assumptions must be made when going from the full
theory of elasticity to a beam model.  These assumptions
yield different mathematical models for the same system
whose validity often depends on the geometry of the element
and the frequency of excitation \cite{Bottega2014}.  Some of
the classical beam theories as well as the associated key
assumptions are shown in Table \ref{tab:beam_models}.  From
this table it is evident that the most inclusive classical
beam model is the Timoshenko beam.

In contrast to the most commonly used Euler-Bernouli (EB)
beam model, the Timoshenko beam (TB) model
\cite{Timoshenko1922} includes shear and rotary inertia
effects in addition to bending deformation and linear
inertia.  This makes the TB model especially more accurate
for beams with smaller length to diameter ratio or with
high-frequency excitation where these effects are not
negligible and thus using the EB model would yield
inaccurate results \cite{Le2001}.  Although the TB model
gives more accurate results than EB, this improvement in
accuracy is associated with a more complicated analysis.
This is because there are two degrees of freedom intrinsic to the TB
model: lateral translation and cross-section rotation while
the EB model has lateral translation as the only degree of
freedom.  Mathematically, TB consists of two second order
coupled partial differential equations (PDEs)
\cite{Timoshenko1922} while EB consists of one PDE.  The two
TB equations can be combined into a hyperbolic PDE while the
EB equation is a parabolic PDE.  This difference causes the
TB model to have two different sets of expressions for the
natural frequencies and mode shapes, thus leading to a
second spectrum for the TB model
\cite{Traill-Nash1953,Abbas1977,Bhashyam1981}.
% However, the second spectrum of frequencies only exists for limited cases where the frquency equation factorizes such as in the hinged-hinged case \cte{Abbas1997}.
For simple boundary conditions, the TB solutions, natural
frequency equations, and mode shapes were derived in
\cite{huang1961}. %\cite{huang1961,VanRensburg2006}.
However a correction for the clamped-clamped frequency
equations that appeared in \cite{huang1961} was later
published in \cite{Kang2014}.  While analytically these
terms are correct, they generally are not suitable for
numerical computations.  This hinders the analysis of a wide
class of systems since beam modes are still regularly used
in assumed modes analysis. Therefore, achieving an accurate and
numerically stable representation, especially in the much
less studied TB models, will be extremely beneficial.
% free-free TB solution: 
% Other BCs: \cite{Traill-Nash1953}
% although for some specific cases the EB model can be more accurate \textbf{cite the weird case(s)} \textbf{Why is it advantageous? What applications? Add references}

% in 2D, Euler Bernoulli beam theory is similar to Kirchhoff plate theory, Tomoshenko beam theory is similar to Reissner-Mindlin plate theory

% The Timoshenko (T) equation is hyperbolic
% And the Euler (E) equation is parabolic. this means that the T equation has a finite phase speed and the E equation has an infinite one. This is important when considering wave effects.

Specifically, it was discovered early on
\cite{Young-Felgar1949Table-Normal-Modes} that numerical
evaluation of higher mode shapes of many EB beam modes (such
as those with clamped or free boundary conditions) can be
numerically unstable---in the sense that evaluation of the
standard forms of the equations for those modes involves
substantial floating point error.  Dowell
\cite{Dowell1984Asymptotic-Approximations-to-beam-modes}
provided an asymptotic form for the high order modes that
circumvented the problem of numerical instability.  Shankar
and Keane \cite{Shankar-Keane1995Energy-Flow-Jointed-Beams}
restructured the equations for the free-free EB beam so as
to isolate the problem to a single term consisting of a very
small number multiplying a very large one.  Using this
approach, they were able to numerically evaluate the first
hundred modes for an EB free-free beam.  Tang
\cite{Tang2003Numerical-evaluation-of-uniform-beam-modes}
also rearranged the terms of the mode shapes, but did so in
a way that reduces catastrophic cancellations which are
numerical errors caused by taking the difference of nearly
identical terms.  However, his formulation still involved
taking differences of very large numbers and the core
problem persisted. Van Rensburg et al  \cite{VanRensburg2006}
studied the
Timoshenko beam model for pinned-pinned and cantilever
boundary conditions.  While the expressions they provide for
the modes are numerically unstable, they derive asymptotic
expressions that can be used to approximate the eigenvalues
for the two boundary condition cases they considered.

Gon\c{c}alves et al
\cite{Gonccalves_etal2007Numerical-High-Order-Beam-Modes}
rearranged the mode equations for common EB boundary
conditions in a manner similar to that done by Shankar and
Keane \cite{Shankar-Keane1995Energy-Flow-Jointed-Beams}.
Similar to
\cite{Shankar-Keane1995Energy-Flow-Jointed-Beams}, they
presented exact terms for numerically computing the mode
shapes up to the $200$th term, and provided approximate
terms for modes higher than $200$ but at the cost of losing
some accuracy at the lower modes.  In a recent publication
\cite{Goncalves2018}, Gon\c{c}alves et al presented an exact
formulation for the mode shapes of uniform EB beams subject
to common boundary conditions.  In contrast to their earlier
publication
\cite{Gonccalves_etal2007Numerical-High-Order-Beam-Modes},
the expressions in \cite{Goncalves2018} are exact and
numerically stable for the full frequency range where the EB
beam theory is valid.  The same authors provide a strategy
for numerically stabilizing the characteristic frequency
equation.

% Neither Gon\c{c}alves et al \cite{Gonccalves_etal2007Numerical-High-Order-Beam-Modes,Goncalves2018} nor Shankar and Keane \cite{Shankar-Keane1995Energy-Flow-Jointed-Beams} tackled the problem of the numerical stabilization of the characteristic equation where the terms also grow which makes solving for the eigenvalues more difficult as well.

The issue of numerical stability of modes is even more
prominent for stepped beams, both Euler-Bernoulli and
Timoshenko, where fewer than  $10$ modes
can be predicted reliably
\cite{Xu_etal2014Numerical-Evaluation-High-Order-Stepped-Beam-Modes,Cao2016}.
Xu et
al.~\cite{Xu_etal2014Numerical-Evaluation-High-Order-Stepped-Beam-Modes}
and Cao et al~\cite{Cao2016} have mitigated the issue by
defining a set of local coordinates that decrease the growth
rate of the hyperbolic sine and cosine terms in
Euler-Bernoulli and Timoshenko beams, respectively.
However, their approach does not solve the core problem and
numerical stabilities continue to be a problem in their
formulation.

Therefore, the problem of obtaining closed form expressions
for the mode shapes that are well-conditioned for the whole
span of the beam and throughout the frequency range is still
a current problem especially for the Timoshenko beam model.

While finite element can be used to obtain eigenfrequencies and mode shapes, analytic expressions are still useful for gaining an understanding of the character of the solution, particularly asymptotic behavior. 
Additionally, analytic expressions (when they can be evaluated) are necessary to gain confidence in numerical methods such as finite elements (mesh convergence, code validation, etc.).
Therefore, closed form and numerically stable solutions for the eigenvalue problem of beams comprise an important analysis tool.

Further, for Timoshenko beams numerical instability occurs at the lower modes which get singular. 
Therefore, even using a dozen of modes for analyzing the vibrations of these beams can be problematic.
On the other hand, for the Euler-Bernoulli beam the number of singular modes depends on the problem, while how many of these modes one needs depends on the application.
 Civil engineering applications often utilize lower modes; however, mechanical engineering applications often require considering higher modes. 
 For instance, if one is using the assumed modes method it is common to utilize several tens of modes. 
 Consequently, the lack of numerically stable expressions can hinder the analysis of Euler-Bernoulli and Timoshenko beams.

In this paper we present an approach for the numerical
stabilization of both the characteristic equation and the
modes of the Timoshenko beam model.  We present the approach
and tabulate the results for a wide range of common boundary
conditions.  For completeness, we also present the results
for the EB model, where our formulation is mathematically
equivalent to the
one presented in \cite{Goncalves2018}.  We are motivated by
both (1) the importance of obtaining numerically stable
expressions for the TB model, and (2) the need for
presenting EB and TB results in one consistent notation and
for all the common boundary conditions in one manuscript.
Figure \ref{fig:common_BCs} shows the boundary conditions
used in this paper.  All the combinations of these boundary
conditions are considered and the corresponding results are
tabulated.

This paper is organized as follows. In Sections
\ref{sec:EB_eom}  and  \ref{Sec:gen_forms} the equations of
motion for the EB and the TB models, respectively, are
non-dimensionalized and the corresponding mode equations are
derived.  The source of
the numerical instability in the characteristic equations
and the mode shape expressions are described
in Section \ref{sec:problem}.
% Section \ref{Sec:num_stab} derives
Numerically stable expressions for both EB
(Section \ref{Sec:EB_eig_forms}) and TB (Section
\ref{sec:TB_regularization}) models, including some useful
identities
are presented in Section \ref{Sec:num_stab}. % in Section \ref{sec:usefulIDs}
(An expanded discussion of the pinned-pinned case
of  a TB is presented in \ref{Sec:TB_pinned_pinned}.)
Section \ref{Sec:Results} shows the results obtained using the
well-conditioned expressions.  Section \ref{sec:Verification-Outline}
 outlines the use of finite element
analysis to verify the exact solutions derived above and
a more complete discussion of the finite element analysis and
the verification process is presented in \ref{sec:Verification}.
The paper
ends with conclusions in Section \ref{Sec:conclusions}.  Two additional
appendices are also provided:
\ref{sec:Apdx-Derive-Modes} provides more details on mode
derivations, while \ref{sec:roller_BCs} includes
the numerically stable expressions for all the roller
boundary condition combinations.

In the following, the nomenclature of \cite{VanRensburg2006}
is used 
with the exception that the shear coefficient is represented
as $\kappa$ rather than $\kappa^2$.  The authors of this monograph
performed their derivations for exact
and numerically stable solutions for Euler Bernoulli and Timoshenko
beams before \cite{Goncalves2018} was published, so the following
will employ the formalism of the authors' derivation.

%------------------------------
\begin{table}[htbp]
\centering
	\begin{tabular}{c|cccc}
	 								& \thead{Shear\\ deformation} 	& \thead{Bending \\ deformation}  	& \thead{Linear\\ inertia}	& \thead{Rotary\\ inertia} \\
	 								\toprule
	\thead{Euler-Bernoulli} 		& \xmark						& \cmark							& \cmark				&  \xmark \\
	\thead{Rayleigh} 				& \xmark						& \cmark							& \cmark				& \cmark \\
	\thead{Shear} 					& \cmark						& \xmark							& \cmark				& \xmark \\
	\thead{Timoshenko} 				& \cmark						& \cmark							& \cmark				& \cmark  \\ 
	\end{tabular}
	\caption{Beam models. A check mark in any column
          means that the factor stated in the header of the
          column is included in the model, while a cross
          means that the effect of that factor is
          neglected.}
	\label{tab:beam_models}
\end{table}
\pdfoutput=1
%******************************************
\section{The general eigenvalue problem for the Euler-Bernoulli beam}
\label{sec:EB_eom}
%******************************************
The lateral vibrations $w$ of a uniform Euler-Bernoulli beam are governed by the partial differential equation 
%----------------------------------------
\begin{equation} \label{eq:EB1}
  EI \frac{\partial^4}{\partial x^4} w
  = -\rho A \frac{\partial^2}{\partial t^2} w,
\end{equation}
%----------------------------------------
where $\rho$ is the density, $A$ is the area of
cross-section, $E$ is Young's modulus, and $I$ is the area
moment of inertia of the cross section about the neutral
axis.  The beam has length $L$, and we use the
non-dimensional parameters
%----------------------------------
\begin{center}
  $\tau=t/T_{\rm E}$, \quad $\zeta =x/L$
\end{center}
\noindent where $T_{\rm E}=L\sqrt{\rho/E}$.
\noindent The dimensionless displacement is
\begin{center}
 $\tilde{w}(\zeta, \tau)=w(\zeta L, \tau T_{\rm E})/L$
\end{center}
%----------------------------------

\begin{figure}[htbp]
\centering
\includegraphics[width=0.8\textwidth]{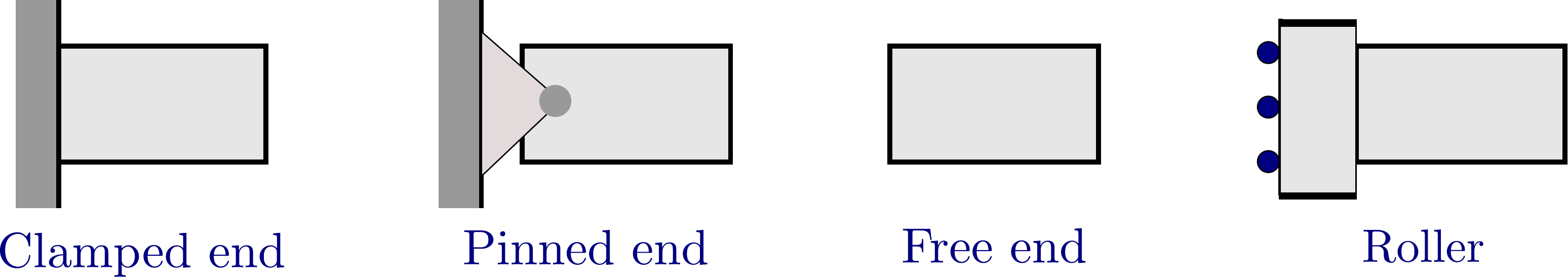}
\caption{Common beam boundary conditions.}
\label{fig:common_BCs}
\end{figure}

%% , and we define the dimensionless quantities \\
%% \noindent$\zeta = x/L, \quad \tau = t/T_{\rm E}, \quad \alpha=AL^2/I, \quad 
%% \rm{ and }\quad \tilde{w}(\zeta, \tau)=w(\zeta L, \tau T_{\rm E})/L$ which
These can be used to derive the following expressions
%----------------------------------------
\begin{equation}
  \frac{\partial^4 w}{\partial x^4} =
  (1/L^3) \frac{\partial^4 \tilde{w} }{\partial \zeta^4}
  \quad \mbox{and} \quad
  \frac{\partial^2 w}{\partial t^2} 
  =(L/T_{\rm E}^2) \frac{\partial^2 \tilde{w}}{\partial \tau^2}.
\end{equation}
%----------------------------------------
Substituting the above into Eq.~\eqref{eq:EB1} gives
%----------------------------------------
\begin{equation}
  \frac{1}{\alpha} \frac{\partial^4 \tilde{w}}{\partial \zeta^4}
  = - \frac{\partial^2 \tilde{w}}{\partial \tau^2}.
\end{equation}
\noindent where $ \alpha=AL^2/I$.
%----------------------------------------
We appeal to separation of variables:
$\tilde{w}(\zeta,\tau) = U(\zeta)Y(\tau)$ and obtain
%----------------------------------------
\begin{equation}
  \frac{\frac{1}{\alpha} \frac{\partial^4 U}{\partial  \zeta^4} }{U}
  =- \frac{ \frac{\partial^2 Y}{\partial \tau^2} }{Y} = \lambda, \quad \text{where $\lambda$ is assumed positive.}
\end{equation}
\noindent Dimensionless eigenvalue $\lambda$ is related to
physical quantities by $\sqrt{\lambda}  = T_E \hat{\omega}$ where
$\hat{\omega}$ is the physical natural frequency predicted by the model.
%----------------------------------------
We are now interested in the form of $U$, so we examine the equation
%----------------------------------------
\begin{equation}
\label{eq:EB_eigProb}
  \frac{\partial^4 U}{\partial  \zeta^4}
  = \alpha \lambda U.
\end{equation}
%----------------------------------------
Assuming $U = e^{\mu\zeta}$ and substituting it into the above equation gives
%----------------------------------------
\begin{equation}
e^{\mu\zeta} \left(\mu^4 - \alpha \lambda\right) = 0, \quad \text{ so } \quad 
  \mu = \left(\alpha \lambda\right)^{\frac{1}{4}}.
\end{equation}
%----------------------------------------
Considering of both real and imaginary roots yields
the following real expressions
for the general form for the Euler-Bernoulli modes
%----------------------------------------
\begin{equation} \label{eq:FullExpansion}
    u(\xi) = A_1(\mu) \sinh(\mu \xi)
    +A_2(\mu) \cosh(\mu \xi)
    +A_3(\mu) \sin(\mu \xi)
    +A_4(\mu) \cos(\mu \xi),
\end{equation}
%----------------------------------------
where the constants $A_1$ through $A_4$ depend on the
boundary conditions.  The most common boundary conditions
for the Euler-Bernoulli beam are similar to the ones shown
in Fig.~\ref{fig:common_BCs}, and the corresponding
mathematical expressions are listed in Table
\ref{tab:EBFourStandardBoundaryConditions}.
%----------------------------------
\begin{table}[h!]
  \begin{center}
  \begin{tabular}{ccc}
    \toprule
    Condition Name & Equation 1 & Equation 2 \\
    \toprule
    Cantilever        & $u=0$              & $u^{\prime}=0$           \\
    Simply Supported  & $u=0$              & $u^{\prime\prime}=0$      \\
    Free              & $u^{\prime\prime} =0$ & $u^{\prime\prime\prime}=0$ \\
    Roller            & $u^\prime =0$       & $u^{\prime\prime\prime}=0$ \\
    \bottomrule
  \end{tabular}
  \end{center}
  \caption{The four standard boundary conditions for the Euler-Bernoulli beam.
    \label{tab:EBFourStandardBoundaryConditions}
    }
\end{table}
%------------------------------
%------------------------------
%------------------------------

%!TEX root = ../Timoshenko.tex
\pdfoutput=1
\section{The general eigenvalue problem for the Timoshenko beam} \label{Sec:gen_forms}
The Timoshenko beam model is given by the two coupled partial differential equations \cite{VanRensburg2006}
%----------------------------------
\begin{align}
\label{eq:TB_eom}
\rho A \frac{\partial ^2 u}{\partial t^2} &=\frac{\partial (AG \kappa (\frac{\partial u}{\partial x}-\phi))}{\partial x},\\
\rho I \frac{\partial ^2\phi}{\partial t^2} &= AG \kappa  (\frac{\partial u}{\partial x}-\phi)+\frac{\partial(EI\frac{\partial \phi}{\partial x})}{\partial x},
\end{align}
%----------------------------------
where $u$ is the lateral deflection, $\phi$ is the angle of rotation of the beam's cross-section, $A$, $E$ and $I$ have the same meaning described in Section \ref{sec:EB_eom}, $\kappa$ is the shear coefficient which corrects for the variation in shear along the cross section due to shear deformation in the Timoshenko beam model \cite{Hutchinson2001}, while $G$ is the shear modulus.
In order to derive these equations, and incorporate the appropriate boundary conditions, the following expressions for the bending moment and the shear force were used
%----------------------------------
\begin{equation} 
\label{eq:MomentandShear}
M = EI\frac{\partial \phi}{\partial x}, \quad
V = AG \kappa  (\frac{\partial u}{\partial x}-\phi). 
\end{equation}  
%----------------------------------

We now define the dimensionless parameters
$u^*(\zeta, \tau)=u(\zeta L, \tau T_{\rm T})/L,$
\quad and \quad  $T_{\rm T}=L\sqrt{\frac{\rho}{G\kappa}}$, \quad 
We also define the following constants\\
\begin{center}
$\alpha=\frac{AL^2}{I}$, $\beta=\frac{AG\kappa L^2}{EI}$ and $\gamma=\frac{\beta}{\alpha}$.\\
\end{center}

Utilizing these non-dimensional parameters, the Timoshenko
beam model is written as
%----------------------------------
\begin{align}
  \frac{\partial ^2 u^*}{\partial \tau ^2} &= \frac{\partial
    ^2 u^*}{\partial \zeta^2}-\frac{\partial
    \phi^*}{\partial \zeta} ,\\
  \frac{\partial ^2 \phi^*}{\partial \tau^2} &=
  \frac{1}{\gamma}\frac{\partial^2 \phi^*}{\partial \zeta^2} +
  \alpha \frac{\partial u^*}{\partial \zeta}-\alpha\phi^* .
\end{align}
%----------------------------------
Following the approach described in \cite{VanRensburg2006},
a separation of variables of the form
$u^*(\zeta,\tau)=Y^*(\tau)U^*(\zeta)$ and
$\phi^*(\zeta,\tau)=Y^*(\tau)\Phi^*(\zeta)$ is considered
and the resulting eigenvalue problem is given by
%----------------------------------
\begin{equation}
\begin{aligned}
-U^{*''} + \Phi^{*'} &= \lambda U^*,\\
-\frac{1}{\gamma}\Phi^{*''} - \alpha U^{*'} + \alpha \Phi^* &= \lambda \Phi^* 
\end{aligned}\label{eq: nd eigenvalue}
\end{equation}
%----------------------------------
where $\lambda$ is the non-negative eigenvalue and 
$\begin{bmatrix}
U^*\\
\Phi^*
\end{bmatrix}$
is the eigenfunction. Here $\frac{Y^{*''}}{Y^*}=-\lambda$
and $\sqrt{\lambda}$ is the non-dimensional natural
frequency $\sqrt{\lambda}=\hat{\omega} T_T$, where
$\hat{\omega}$ is the physical frequency predicted by the model. To solve the
eigenvalue problem, the eigenfunctions are assumed to be of
the form $e^{m\zeta}\mathbf{w}$ resulting in the following
equation
%----------------------------------
\begin{equation} 
\begin{bmatrix}
-m^2 -\lambda & m \\
-\alpha m & -\frac{1}{\gamma}m^2 +(\alpha - \lambda)
\end{bmatrix}
\begin{bmatrix}
w_{1}\\
w_{2}
\end{bmatrix}
=\begin{bmatrix}
0 \\
0
\end{bmatrix}.\label{eq: eigenvalue eigenfn eqn}
\end{equation}
%----------------------------------
Equating the determinant to zero for non-trivial solutions,
i.e., $m^4 +\lambda(1+\gamma)m^2
+\gamma\lambda(\lambda\alpha)=0$ leads to following roots of
$m$
%----------------------------------
\begin{equation}
\label{eq:Delta}
m^2 = -\frac{1}{2}\lambda(1+\gamma)(1\pm\Delta^\frac{1}{2}),  \quad
\text{ where } 
\Delta=1-\frac{4\gamma}{(1+\gamma)^2}(1-\frac{\alpha
}{\lambda}).
\end{equation}
%----------------------------------
The parameter $\Delta$ in Eq.~\eqref{eq:Delta} is always
greater than $0$ but can be greater than, less than or equal
to $1$, depending upon whether $\lambda<\alpha$,
$\lambda>\alpha$ or $\lambda=\alpha$.  The three cases lead
to different expressions for the mode shapes which are
presented in Sections
\ref{sec:lamLessAlph}--\ref{sec:lamGrAlph}
\cite{VanRensburg2006}. 
\ref{sec:Apdx-Derive-Modes} provides some discussion on the
derivation of Eqs.~\eqref{eq: general modeshape
  lambda<alpha}, \eqref{eq: general modeshape lambda=alpha},
and \eqref{eq: general modeshape lambda>alpha}.
%**********************************
\subsection{The case \boldsymbol{$\lambda<\alpha$}}
\label{sec:lamLessAlph}
%**********************************
%
In this case $m$ has $2$ real roots denoted by $\pm \mu$ and
$2$ imaginary roots denoted by $\pm \omega i$.  The
eigenfunctions using Eq.~\eqref{eq: eigenvalue eigenfn eqn}
are then expressed by
%
%----------------------------------
\begin{equation}
\label{eq: general modeshape lambda<alpha}
\begin{bmatrix}
U^*(\zeta)\\
\Phi^*(\zeta)
\end{bmatrix}
=A_1\begin{bmatrix}
\sinh \mu \zeta \\
\frac{\lambda + \mu ^2}{\mu}\cosh\mu \zeta
\end{bmatrix}
+A_2\begin{bmatrix}
\cosh\mu\zeta \\
\frac{\lambda + \mu ^2}{\mu}\sinh\mu \zeta
\end{bmatrix}
+A_3\begin{bmatrix}
\sin\omega\zeta \\
-\frac{\lambda - \omega ^2}{\omega}\cos\omega \zeta
\end{bmatrix}\\
+A_4\begin{bmatrix}
\cos\omega\zeta \\
\frac{\lambda - \omega ^2}{\omega}\sin\omega \zeta
\end{bmatrix}.
\end{equation} 
%----------------------------------

%**********************************
\subsection{The case \boldsymbol{$\lambda=\alpha$}:}
\label{sec:lamEqAlph}
%**********************************
The roots in this case are two imaginary values denoted by
$\pm \omega i$ and $0$ with multiplicity $2$.  The
eigenfunctions are then given by
%----------------------------------
\begin{equation}
\begin{bmatrix}
U^*(\zeta)\\
\Phi^*(\zeta)
\end{bmatrix}
=A_1\begin{bmatrix}
0 \\
1
\end{bmatrix}
+A_2\begin{bmatrix}
1 \\
\alpha \zeta
\end{bmatrix}
+A_3\begin{bmatrix}
\sin\omega\zeta \\
-\frac{\lambda - \omega ^2}{\omega}\cos\omega \zeta
\end{bmatrix}
+A_4\begin{bmatrix}
\cos\omega\zeta \\
\frac{\lambda - \omega ^2}{\omega}\sin\omega \zeta
\end{bmatrix}.\label{eq: general modeshape lambda=alpha}
\end{equation}
%----------------------------------
%**********************************
\subsection{The case \boldsymbol{$\lambda>\alpha$}}
\label{sec:lamGrAlph}
%**********************************
All the $4$ roots are imaginary denoted by $\pm \theta i$
and $\pm \omega i$. The eigenfunctions are then given by
%----------------------------------
\begin{equation}
\begin{bmatrix}
U^*(\zeta)\\
\Phi^*(\zeta)
\end{bmatrix}
=A_1\begin{bmatrix}
\sin \theta \zeta \\
-\frac{\lambda - \theta ^2}{\theta}\cos\theta \zeta
\end{bmatrix}
+A_2\begin{bmatrix}
\cos\theta\zeta \\
\frac{\lambda - \theta ^2}{\theta}\sin\theta \zeta
\end{bmatrix}
+A_3\begin{bmatrix}
\sin\omega\zeta \\
-\frac{\lambda - \omega ^2}{\omega}\cos\omega \zeta
\end{bmatrix}\\
+A_4\begin{bmatrix}
\cos\omega\zeta \\
\frac{\lambda - \omega ^2}{\omega}\sin\omega \zeta
\end{bmatrix}.\label{eq: general modeshape lambda>alpha}
\end{equation}
%----------------------------------
%

The real quantities $\omega$, $\mu$ (for the case
$\lambda<\alpha$) and $\theta$ (for the case
$\lambda>\alpha$) are given in terms of $\lambda$ as
\cite{VanRensburg2006}
%----------------------------------
\begin{equation}
\omega ^2 = \frac{1}{2}\lambda (1+\gamma)(\Delta ^\frac{1}{2} +1), \quad
\mu ^2 = \frac{1}{2}\lambda (1+\gamma)(\Delta ^\frac{1}{2} -1), \quad
\theta ^2 = \frac{1}{2}\lambda (1+\gamma)(1-\Delta ^\frac{1}{2}).
\end{equation}
%----------------------------------
The values of the coefficients $A_1$ through $A_4$ along
with the quantity $\lambda$ which determines the natural
frequencies depend on the specific boundary conditions and
can be found by imposing the boundary conditions on the
above expressions for mode shape.  Figure
\ref{fig:common_BCs} shows the most common boundary
conditions, and Table
\ref{tab:FourStandardBoundaryConditions} lists the
corresponding mathematical expressions.
% The free-free boundary condition is further discussed in Section \ref{sec:free_free_TB}.
%----------------------------------
\begin{table}[h!]
  \begin{center}
  \begin{tabular}{ccc}
    \toprule
    Condition Name & Equation 1 & Equation 2\\
    \toprule
    Cantilever        & $u=0$             & $\phi=0$       \\
    Simply Supported  & $u=0$             & $\phi^\prime=0$ \\
    Free              & $u^\prime -\phi=0$ & $\phi^\prime=0$ \\
    Roller            & $u^\prime -\phi=0$ & $\phi=0$        \\
    \bottomrule
  \end{tabular}
  \end{center}
  \caption{The four standard boundary conditions for the Timoshenko beam.
    \label{tab:FourStandardBoundaryConditions}
    }
\end{table}
%------------------------------
%********************************************************
\section{Solution strategy for the eigenvalue problem}
%********************************************************
This section details the strategy % that we
devised for
solving the eigenvalue problem.
Symbolic Matlab\textsuperscript{\textregistered} was used, as
well as Mathematica\textsuperscript{\textregistered}
to perform the algebraic manipulations
needed to solve the resulting system of equations.  This
section assumes that $\lambda >0$, i.e., we are not
concerned with rigid body modes at $\lambda=0$ or
eigenfrequencies with multiplicity 2 \cite{geist1997double}
(see Section \ref{sec:doubleEig}).

The process starts by imposing the boundary conditions from
Table \ref{tab:FourStandardBoundaryConditions} for the beam
case under investigation, and using symbolic variables, we
can write down a system of equations
%---------------------------------------
\begin{equation}
\mathbf{B} \, \mathbf{a} = \mathbf{0},
\end{equation}
%---------------------------------------
where the the entries of matrix $\mathbf{B}$ are
$b_{ij}=b_{ij}(\mu(\lambda), \omega(\lambda))$, where $i,j
\in \{1, 2, 3 ,4\}$, are obtained by applying the boundary
conditions, and the entries of vector $\mathbf{a}$ are the
coefficients $A_i$, where $i\in \{1, 2, 3, 4 \}$.  It is
important to first check the rank of the matrix
$\mathbf{B}$.  If $\mathbf{B}$ is full rank, when the
following process will reliably produce the needed
expressions for the eigenmode coefficients.  However, if
$\mathbf{B}$ is rank deficient, then the eigenvalue problem
needs to be studied more closely to identify any additional
constraints on the coefficients $A_1$--$A_4$.  For all but
one of the cases that we have studied the matrices had full
rank.  The only case with a rank deficient $\mathbf{B}$
matrix was the pinned-pinned case with $\lambda=\alpha$.
This case is described in detail in \ref{Sec:TB_pinned_pinned}.

After ensuring that $\mathbf{B}$ has full rank, Gaussian
elimination can then be used by choosing one of the $a_i$s
and solving for the other entries in $\mathbf{a}$ in terms
of this $a_i$.  However, while performing the Gaussian
elimination, it is necessary to perform column pivoting
since otherwise the Gaussian elimination may not be stable
\cite{Trefethn1997}.  Assume that after column pivoting the
linear system reads
%---------------------------------------
\begin{equation}
\label{eq:permutedLinSys}
\tilde{\mathbf{B}} \, \tilde{\mathbf{a}} = \mathbf{0},
\end{equation}
%---------------------------------------
and that we have a bijective permutation map $\sigma$ from
the set $\{1, 2, 3, 4 \}$ onto itself such that
% %---------------------------------------
\begin{equation}
\label{eq:permutVec}
\tilde{a}_i = a_{\sigma(i)} \textrm{ where } i \in \{1, 2,
3, 4\}; \quad \textrm{ and } \quad
\tilde{B}_{i,j}=B_{i,\sigma(j)} \textrm{ where } j \in \{1,
2, 3, 4\}.
\end{equation}
%---------------------------------------
Note that the map $\sigma$ is invertible, i.e., we can write
$\tilde{a}_{\sigma^{-1}(i)} = a_i$ and
$\tilde{B}_{i,\sigma^{-1}(j)}=B_{i,j}$.  We can now use
Gaussian elimination on Eq.~\eqref{eq:permutedLinSys} and
then use Eq.~\eqref{eq:permutVec} along with the inverse map
$\sigma^{-1}$ to solve for the coefficients in
Eqs.~\eqref{eq: general modeshape lambda<alpha}--\eqref{eq:
  general modeshape lambda>alpha}.  Since we can only solve
for three of the coefficients in terms of the fourth, we
choose the term $\tilde{a}_{4}$ as the free coefficient, set
$\tilde{a}_{4}=1$, and find $\tilde{a}_1$--$\tilde{a}_3$ in
terms of it.  This is accomplished by constructing and
symbolically solving (via Gauss elimination) the augmented
matrix
%---------------------------------------
\begin{equation}
  \begin{bmatrix}[ccc:c]
    \tilde{B}_{1, \sigma(1)} & \tilde{B}_{1, \sigma(2)} & \tilde{B}_{1, \sigma(3)} & -\tilde{B}_{1, \sigma(4)} \\
    \tilde{B}_{2, \sigma(1)} & \tilde{B}_{2, \sigma(2)} & \tilde{B}_{2, \sigma(3)} & -\tilde{B}_{2, \sigma(4)} \\
    \tilde{B}_{3, \sigma(1)} & \tilde{B}_{3, \sigma(2)} & \tilde{B}_{3, \sigma(3)} & -\tilde{B}_{3, \sigma(4)}     
  \end{bmatrix}.
\end{equation}
%---------------------------------------

%********************************************************
\subsection{Orthonormalization of the modes}
\label{sec:mode_orthonormal}
%********************************************************
There are several choices for the orthonormalization of the
modes.  One of the useful choices, especially in the method
of assumed modes, is to normalize the modes with respect to
the beam's mass.  For the Euler-Bernoulli beam this is
obtained according to
%-----------------------------------
\begin{equation}
\textrm{Dimensional: } \quad \int\limits_{0}^{L}\rho A u_i
(x) u_j (x)\, dx = \delta_{ij}; \quad
\textrm{Non-dimensional:} \quad \int\limits_0^{1}{\alpha
  U_i(\zeta) U_j(\zeta)\, d\zeta}=\delta_{ij},
\end{equation}
%-----------------------------------
where $\delta_{ij}$ is the Kronecker delta function.

In the case of the Timoshenko beam, one point to emphasize
is that the eigenmodes are vector valued; therefore, any
choice of normalization involves the inner product of two
vector-valued eigenmodes.  To elaborate, the mass
orthonormalization for the Timoshenko model is given by
%-------------------------------------
\begin{subequations}
  \label{eq:orthonormalizn reln for Timoshenko eignfn}
  \begin{align}
    \textrm{Dimensional:} \quad  \int\limits_{0}^{L}{[\rho A U_i (x)U_j (x) + \rho I \Phi^i (x) \Phi ^j (x)] dx} &= \delta_{ij}; \\
    \textrm{Non-dimensional:} \quad \int\limits_{0}^{1}{[U^*_i (x)U^*_j (\zeta) + \frac{1}{\alpha} \Phi^*_i (x) \Phi^*_j (\zeta)]\, d\zeta} &= \delta_{ij}.
  \end{align}
\end{subequations}
%-------------------------------------
where $U$ and $\Phi$ are the dimensional mode shapes and
they are related to the non-dimensional mode shapes
according to $U(x)=L \, U^*(\zeta)$ and $\Phi(x)=\Phi^*
(\zeta)$, respectively.  Note that these normalizations can
be used for both rigid, and non-rigid beam modes.  As an
example, the orthonormal rigid-body modes for a free-free
beam are given by
%-------------------------------------
\begin{equation}
\label{eq: normalized rigid body modes}
\begin{bmatrix}
U(x)\\
\Phi (x)
\end{bmatrix} =\frac{1}{\sqrt{\rho A L}}\begin{bmatrix}
1\\
0
\end{bmatrix}, \quad \mbox{and}
\qquad
\begin{bmatrix}
U(x)\\
\Phi (x)
\end{bmatrix} =\frac{12}{\sqrt{\rho A L^3 + 12\rho I L}}\begin{bmatrix}
\frac{L}{2}-x\\
-1
\end{bmatrix}.
\end{equation}
%-------------------------------------

%********************************************************
\subsection{Double eigenvalues}
\label{sec:doubleEig}
%********************************************************
Although double eigenvalues besides the ones corresponding
to rigid body modes can theoretically occur in beam models
\cite{geist1997double}, their occurrence is generally
unlikely \cite{VanRensburg2006}.  Therefore, while we are
aware of the possibility of double eigenvalues, we do not
consider that case in this manuscript.

% \clearpage  % added temporarily to faciliate identifying contributing
%             .tex files
% the three eigenmodes regimes (already written)
%

%!TEX root = ../Timoshenko.tex
\pdfoutput=1
%******************************************
\section{Problem description}
\label{sec:problem}
%******************************************
The standard equations for the
eigenfrequencies and eigenmodes of beams having certain
boundary conditions such as clamped or free ends can be
numerically ill-posed in that their solution is subject to
large floating point error.  
This can be attributed to the
form of Eqs.~\eqref{eq: general modeshape lambda<alpha} and
\eqref{eq:FullExpansion} which involve differences of terms containing
$\sinh(\mu\xi)$ and $\cosh(\mu \xi)$, that grow unboundedly as $\mu$
increases.  
These issues are manifested in both the
characteristic functions and the mode shapes.

\begin{figure}[htbp]
\begin{center}
\resizebox{\textwidth}{!}{
  \includegraphics{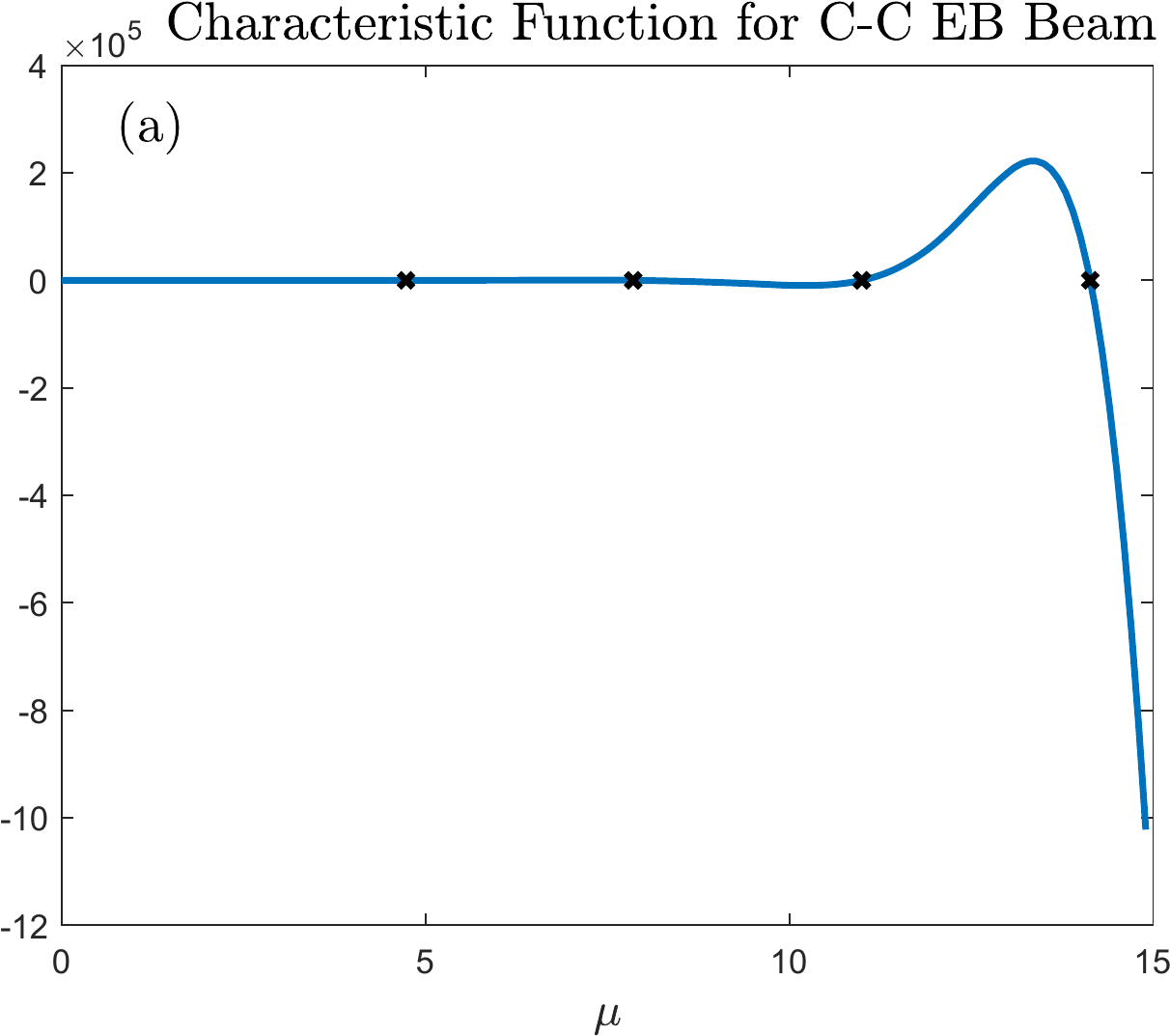}
  \includegraphics{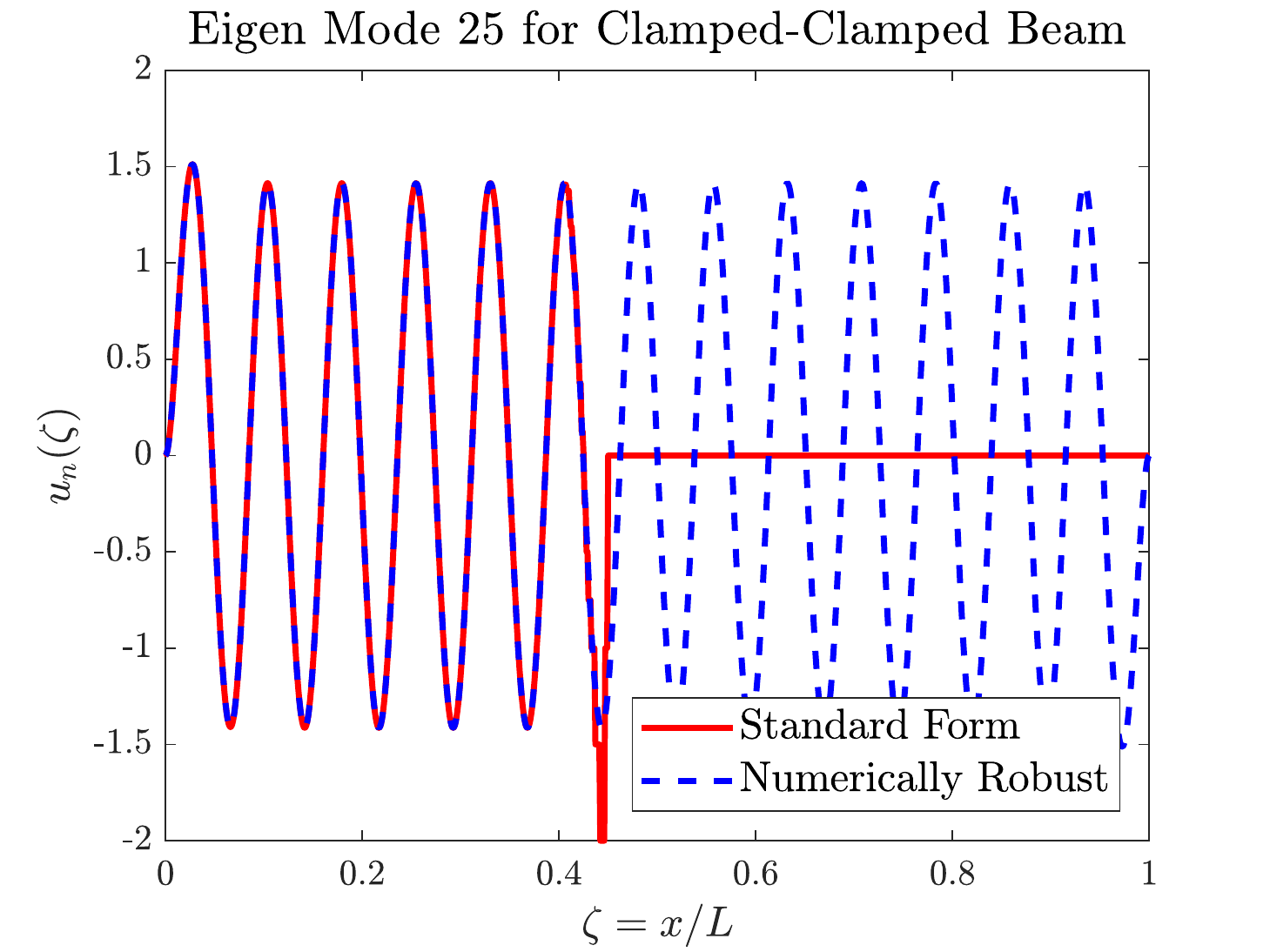}
}
\caption{\em Standard form of the (a) characteristic
  equation, and (b) of the high modes for the clamped-clamped case.
  The exponential growth of the hyperbolic terms leads to numerical numerical instability.
 }
\label{fig:C2}
\end{center}
\end{figure}

To illustrate the problem, we consider a
clamped-clamped Euler-Bernoulli beam model.
Fig \ref{fig:C2}a shows the characteristic function
$(\cos(\mu)\cosh(\mu)-1)$
for the dimensionless argument $\mu<15$.  The crosses
are roots of this function and we see that the
characteristic function appears to grow exponentially 
before the fifth root has been found.  In fact the characteristic function will oscillate between positive and negative regions, but within
an exponentially growing envelope.

Numerical instabilities also affect the normal modes.
Specifically, consider the normal modes for the clamped-clamped beam
\begin{equation*}
u_k(\zeta) = \sinh(\mu_k\zeta) -
\frac{\sinh(\mu_k)-\sin(\mu_k)}{\cosh(\mu_k)-\cos(\mu_k)}\,\cosh(\mu_k\zeta)
-\sin(\mu_k\zeta)+
\frac{\sinh(\mu_k)-\sin(\mu_k)}{\cosh(\mu_k)-\cos(\mu_k)}\,\cos(\mu_k\zeta),
\end{equation*}
where $\mu_k$ is the $k{\mbox{th}}$ root for the
characteristic function.  
This expression shows that for large
frequencies the normal modes involve small differences of
large numbers which is the paradigm of numerical
instability.  
Figure \ref{fig:C2}b illustrates this
difficulty: the dashed lines are the displacements for the
$25{\mbox{th}}$ mode as calculated 
using the numerically stable 
expression for this problem derived below - 
and equivalent to one of \cite{Goncalves2018}.
The solid line is obtained from numerical evaluation of
the analytic expression above.  
Around $\zeta=0.45$ Matlab\textsuperscript{\textregistered}
calculates displacements that are ``NaN" and plots zeros.
Figure \ref{fig:C2} is very similar to Figure 1 of
\cite{Goncalves2018} and is included here for continuity and completeness.

% \begin{figure}[htbp]
% \begin{center}
% \resizebox{10cm}{!}{
%   \includegraphics{HighMode}
% }
% \caption{\em The standard form of the equation for
%   high modes of the clamped-clamped case leads to numerical instability.
%  }
% \label{fig:Roots1}
% \end{center}
% \end{figure}
% %-----------------------------------

% \clearpage  % added temporarily to faciliate identifying contributing
%             .tex files
% with examples enumerating cases and using FF as paradigm trouble

% \input{sections/sec-boundary_conditions}
% % all BCs as combinations of 4 constraints. 2 of the four on each side.
% % use the free-free case as an illustartion for the BCs.
% % show the form, the charactersitic equstions, etcetera, and walk through it step-by-step.

% \input{sections/sec-Timoshenko_modes}  % this is the free-free case for Timoshenko

%!TEX root = ../Timoshenko.tex
\pdfoutput=1
\section{Numerical stability and regularization} \label{Sec:num_stab}
This section describes the numerical stabilization of the
characteristic equation and the eigenmodes for the
Euler-Bernoulli (Section \ref{Sec:EB_eig_forms}) and
Timoshenko (Section \ref{sec:TB_regularization}) beam
models.  Before the numerical stabilization method is
described, Section \ref{sec:usefulIDs} lists some useful
identities that are used in the regularization process.

% \clearpage % Inserted temporarily to identify the .tex files associated
%            with given text

% \input{sections/sec-TB_free_free_derivation}
% Are we not presenting the free-free case?
% \input{sections/sec-TB_pinned_pinned}

% Timoshenko
% a full section, but a short section.
% we will discuss the free-free problem since it has the sinh and cosh in it.
%!TEX root = ../Timoshenko.tex
\pdfoutput=1
%---------------------------------------------
\subsection{Some useful identities}\label{sec:usefulIDs}

Depending on the circumstances, it will be helpful to
divide some equations %through
either by
$\cosh(\mu)$ or by $\sqrt{\sinh(\mu)^2+\cosh(\mu)^2}$, so it is useful to
observe that
%----------------------------------------
\begin{equation}
  %% 1/\sinh(\mu) = \frac{2\,{\mathrm{e}}^{-\mu }}{1-{\mathrm{e}}^{-2\,\mu }}
  %% \quad \mbox{and} \quad
  1/\cosh(\mu) = \frac{2\,{\mathrm{e}}^{-\mu }}{1+{\mathrm{e}}^{-2\,\mu }}
   \rightarrow 0 \quad \mbox{as} \quad \mu \rightarrow \infty,
\end{equation}
and define
\begin{equation}
f(\mu) \doteq  \frac{2\,{\mathrm{e}}^{-\mu }}{1+{\mathrm{e}}^{-2\,\mu }}.
\end{equation}

Similarly, we observe that
\begin{equation}
  \frac{1}{\sqrt{\sinh(\mu)^2+\cosh(\mu)^2}}
  = \frac{\sqrt{2}e^{-\mu}}{\sqrt{1+e^{-4\mu}}}
   \rightarrow 0 \quad \mbox{as} \quad \mu \rightarrow \infty,
\end{equation}
\noindent and define
\begin{equation}
  g(\mu) \doteq \frac{\sqrt{2}e^{-\mu}}{\sqrt{1+e^{-4\mu}}}.
\end{equation}
%% \noindent and that
We also observe that 
\begin{equation} \label{eq:SinhRatio}
  \frac{\sinh(\mu)}{\sqrt{\cosh^2(\mu)+\sinh^2(\mu)}}
  =\frac{\sqrt{2}\,\left({1-\mathrm{e}}^{-2\,\mu }\right)}
  {2\,\sqrt{1+{\mathrm{e}}^{-4\,\mu }}}
  \rightarrow \frac{\sqrt{2}}{2} \quad \mbox{as} \quad \mu \rightarrow \infty,
\end{equation}
\noindent and
\begin{equation}\label{eq:CoshRatio}
  \frac{\cosh(\mu)}{ \sqrt{
      \cosh^2(\mu)+\sinh^2(\mu)
  }} =\frac{\sqrt{2}\,\left(1+{\mathrm{e}}^{-2\,\mu }\right)}
       {2\,\sqrt{1+{\mathrm{e}}^{-4\,\mu }}}
  \rightarrow \frac{\sqrt{2}}{2} \quad \mbox{as} \quad \mu \rightarrow \infty.
\end{equation}
%----------------------------------------

It is now useful to define angle $\psi$ such that
%----------------------------------------
\begin{equation}
\label{eq:psi_def}
  \psi(\mu) \doteq \atantwo ({1-\mathrm{e}}^{-2\,\mu }, {1+\mathrm{e}}^{-2\,\mu })
    \rightarrow \frac{\pi}{4} \quad \mbox{as} \quad \mu \rightarrow \infty.
\end{equation}
%----------------------------------------
We can now write Eqs.~\eqref{eq:SinhRatio} and \eqref{eq:CoshRatio} as
%----------------------------------------
\begin{equation} \label{eq:SinhRatio2}
  \frac{\sinh(\mu)}{\sqrt{\cosh^2(\mu)+\sinh^2(\mu)}}
  = \sin(\psi(\mu)),
  % \rightarrow \frac{1}{\sqrt{2}} \quad \mbox{as} \quad \mu \rightarrow \infty
\end{equation}
%----------------------------------------
and
%----------------------------------------
\begin{equation}\label{eq:CoshRatio2}
  \frac{\cosh(\mu)}{ \sqrt{
      \cosh^2(\mu)+\sinh^2(\mu)
  }} = \cos(\psi(\mu)).
 % \rightarrow \frac{1}{\sqrt{2}} \quad \mbox{as} \quad \mu \rightarrow \infty
\end{equation}
%----------------------------------------

These identities are used to provide the numerically tractable forms 
for characteristic functions  presented below.
%that are 
%provided in the third column of Table \ref{tab:ScaledCharEq}.

Two other expressions that are useful in expanding the eigenmodes
are
\begin{equation} \label{eq:HyperbolicRatios}
  \frac{\sinh(\mu\zeta)}{\cosh(\mu)}
  = \frac{e^{-\mu(1-\zeta)}-e^{-\mu(1+\zeta)}}
  {1+e^{-2\mu}}, \quad \quad
  \mbox{and} \quad \quad
  \frac{\cosh(\mu\zeta)}{\cosh(\mu)}
  = \frac{e^{-\mu(1-\zeta)}+e^{-\mu(1+\zeta)}}{1+e^{-2\mu}}.
\end{equation}

\subsection{Numerically stable expressions for the Euler-Bernouli beam}
\label{Sec:EB_eig_forms}

The formulas for Euler-Bernouli beam modes have the general
form shown in Eq.~\eqref{eq:FullExpansion}.  A problem of
numerical stability occurs when we consider cases of large
$\mu$ (high frequency) as both $\sinh(\mu \xi)$ and
$\cosh(\mu \xi)$ increase unboundedly.  There has been some
literature on how to address such difficulties and some good
approximate methods have been developed over time.  Here we
present exact equivalent expressions that do not suffer from
the numerical problems.  As mentioned above, these
stabilized expressions for the Euler Bernoulli modes, are
mathematically equivalent to the expressions of
\cite{Goncalves2018}, and are presented here so that the
reader may compare the EB modes with the TB modes using the
same nomenclature.

Of course these difficulties occur only where $A_1$ or $A_2$
are nonzero.  We first consider cases where both of these
coefficients are nonzero in Section \ref{sec:A1A2_notZero},
while the case when these coefficients are zero is described
in Section \ref{sec:A1A2_areZero}.
%*******************************************
\subsubsection{Case 1: non-zero $A_1$ and $A_2$}
\label{sec:A1A2_notZero}
%*******************************************
For these problems, we normalize by the first coefficient so
that $A_1=1$.  In order for the full expression of Equation
\ref{eq:FullExpansion} to stay bounded, the exponential
growth of the $\sinh(\mu \xi)$ and $\cosh(\mu \xi)$ must
exactly cancel.  This motivates us to express
 %----------------------------------------
\begin{equation}
    \frac{A_2}{A_1} = A_2 = - \left(1+P(\mu) e^{-\mu}\right).
\end{equation}
%----------------------------------------
So
%-------------------------------------
\begin{equation}
  \begin{split}
  \sinh(\mu\zeta)
  +A_2 \cosh(\mu \zeta)
  &=
  \sinh(\mu\zeta)
  - \left(1+P(\mu) e^{-\mu}\right)\cosh(\mu \zeta)
  \quad \quad \quad \quad \quad 
  \\
  &=-{\mathrm{e}}^{-\mu \,\zeta }
  -\frac{P(\mu)}{2}\, \left(
    e^{-\mu \,\left(1-\zeta\right)}
  + e^{-\mu \,\left(1+\zeta\right)}
  \right),
  \end{split}
\end{equation}
%-------------------------------------
which is well behaved as long as $P$ is a bounded function
of $\mu$.  In the type of problems discussed here $A_4$ is
generally proportional to $A_2$ ($A_4=R\,A_2$) so we express
%-------------------------------------
\begin{equation}
u(\zeta) = -e^{-\mu \,\zeta }
-\frac{P(\mu)}{2}\, \left(
  e^{-\mu \,\left(1-\zeta\right)}
+ e^{-\mu \,\left(1+\zeta\right)}
\right)
 + A_3(\mu) \sin(\omega \zeta)
    -R(\mu) \left(1+P(\mu) e^{-\mu}\right) \cos(\omega \zeta).
\end{equation}
%-------------------------------------
Numerically stable modal expressions are presented in Table
\ref{tab:StabalizedCases}.  For completeness, the normal
modes for less-singular configurations are also included in
the table, while the roller case can be found in in Table
\ref{tab:rollerStabalizedCases} of \ref{sec:roller_BCs}.

The standard form for the characteristic equations for these
beam configurations is provided in the second column of
Table \ref{tab:ScaledCharEq}.  We see that in several cases,
terms in the characteristic equations grow unboundedly with
their arguement $\mu$.  This also can lead to numerical
difficulty; we would much prefer that the characteristic
function (the left hand side of the characteristic equation)
oscillates between bounded values, such as plus and minus 1.
The characteristic equations are made numerically tractable
by dividing through by $\cosh(\mu)$ or $\sqrt{\cosh^2(\mu) +
  \sinh^2(\mu)}$ as appropriate and employing the identities
and definitions provided in Subsection \ref{sec:usefulIDs}.
This leads to the numerically stable, scaled expressions
shown in Table \ref{tab:ScaledCharEq}.  The last column in
the table shows the asymptotic values for the modes for
large $n$.

%*******************************************
\subsubsection{Case 2: zero $A_1$ or $A_2$}
\label{sec:A1A2_areZero}
%*******************************************
We now consider the case where % $\lambda < \alpha$, and
$A_1=0$ or $A_2=0$, which occurs for the pinned-free and
pinned-pinned beams in Table \ref{tab:TBStabalizedCases},
and all but the clamped roller cases in Table
\ref{tab:rollerStabalizedCases}.

For these cases, the characteristic functions are stabilized
as described in Section \ref{sec:A1A2_notZero}.  However, we
use a different normalization procedure and slightly modify
the stabilization strategy for the modes in comparison to
the cases considered in Section \ref{sec:A1A2_notZero}.
Specifically, the normalization is performed according to
algorithm~\ref{alg:modeNormalization}.
%-------------------------------------------
\begin{algorithm}
\begin{minipage}{0.75\textwidth}
\caption{}\label{alg:modeNormalization}
  \begin{algorithmic}[1]
    \Procedure{Mode normalization for $\lambda < \alpha$}{}
      \If {$A_1 \neq 0$ and $A_2 \neq 0$}
      \State normalize by $A_1$
      \ElsIf  {$A_1 \neq 0$ and $A_2 = 0$}
      \State normalize by the next non-zero coefficient
      \ElsIf {$A_1= A_3=0$ and $A_2 \neq 0$ and $A_4 \neq 0$}
      \State normalize by $A_4$
      \Else 
      \State normalize by the first non-zero coefficient
      \EndIf     
    \EndProcedure
  \end{algorithmic}
\end{minipage}
\end{algorithm}
%-------------------------------------------

Upon normalizing the modes, we substitute their values into
Eq.~\eqref{eq:FullExpansion}
%% or Eq.~\eqref{eq: general
%%   modeshape lambda<alpha} in the case of the Timoshenko
%% beam,
and then we can use the equations described in Section
\ref{sec:usefulIDs} to numerically stabilize the modes.

\begin{table}[h]
	\centering
    \begin{tabular}{lccc}
    \toprule
    Problem Definition & 
    $P$ &
    $A_3$ & $R$ \\
    \toprule
    Clamped-Clamped & 
    $\frac{2\,\sqrt{2}\,\cos\left(\mu +\frac{\pi }{4}\right)
      -2\,{\mathrm{e}}^{-\mu }}
    {{\mathrm{e}}^{-2\,\mu }
      -2\,{\mathrm{e}}^{-\mu }
      \,\cos\left(\mu \right)+1}$ 
    & $-1$ & $-1$
    %% &
    %% $\cos\left(\mu \right)\,\mathrm{cosh}
    %% \left(\mu \right)-1 = 0$
    \\  \midrule
    Clamped-Free & 
    $-\frac{2\,{\mathrm{e}}^{-\mu }+2\,\sqrt{2}\,\sin\left(\mu +\frac{\pi }{4}\right)}
    {{\mathrm{e}}^{-2\,\mu }
      +2\,{\mathrm{e}}^{-\mu } \,\sin\left(\mu \right)-1}$
    & $-1$ & $-1$
    %% &
    %% $\cos\left(\mu \right)\,\mathrm{cosh}\left(\mu \right)+1=0$
    \\     \midrule
    Clamped-Pinned & 
    $\frac{2\,\sqrt{2}\,\cos\left(\mu +\frac{\pi }{4}\right)
      -2\,{\mathrm{e}}^{-\mu }}
    {{\mathrm{e}}^{-2\,\mu }
      -2\,{\mathrm{e}}^{-\mu }\,
      \cos\left(\mu \right)+1}$
    & $-1$ & $-1$
    %% & 
    %% $\mathrm{cosh}\left(\mu \right)\,\sin\left(\mu \right)
    %% -\,\cos\left(\mu \right)\,\mathrm{sinh}\left(\mu \right) = 0$
    \\     \midrule
    Free-Free & 
    $\frac{2\,{\mathrm{e}}^{-\mu }
      -2\,\sqrt{2}\,\sin\left(\mu +\frac{\pi }{4}\right)}
    {-{\mathrm{e}}^{-2\,\mu }
      +2\,{\mathrm{e}}^{-\mu }
      \,\sin\left(\mu \right)+1}$ & $\,\,\,\, 1$ & $\,\,\,\, 1$
    %% &
    %% $\cos\left(\mu \right)\,\mathrm{cosh}\left(\mu \right)-1=0$
    \\     \midrule
    Pinned-Free & 
	\multicolumn{3}{c}
    {$u(\xi)=\cos(\mu) \left( 
    \frac{\,{\mathrm{e}}^{-\mu (1-\xi)}\,
    -\,{\mathrm{e}}^{-\mu(1+\xi) }} {{\mathrm{e}}^{-2\,\mu }+1}\right)
    + \sin(\mu \xi)$}
    %% & $\cos\left(\mu \right)\,\mathrm{sinh}\left(\mu \right)
    %% -\,\mathrm{cosh}\left(\mu \right)\,\sin\left(\mu \right)=0$
    \\     \midrule
    Pinned-Pinned &  
	\multicolumn{3}{c}
	{$u(\xi)=\sin(\mu \xi)$}
    %% &   $\sin(\mu)=0$
    \\     \bottomrule
 \end{tabular}
    \caption{Euler-Bernouli: Stabilized cases. Note that pinned-free/pinned cases do not need stabilization.}
     \label{tab:StabalizedCases}
\end{table}
%----------------------------------------

%----------------------------------------
\begin{table}[htbp]
\centering
  \begin{tabular}{p{3.2cm}p{4cm}p{3.2cm}p{2.0cm}}
    \toprule
    %% \multirow{2}{*}{\parbox{3.5cm}{Problem Definition \\
    %%   Code}}
    Problem Definition
    & Convent. Char. Eq. & Scaled Char. Eq & Asymp.
    \\   \toprule
    Clamped-Clamped &    $\cos\left(\mu \right)\,\mathrm{cosh}
    \left(\mu \right)-1 = 0$ & $\cos(\mu) - f(\mu)=0$
    & $(2n+1)\pi/2 $
    \\ \midrule
    Clamped-Free &
    $\cos\left(\mu \right)\,\mathrm{cosh}\left(\mu \right)+1=0$
    & $\cos(\mu) + f(\mu) =0 $
    &  $(2n+1)\pi/2 $
    \\ \midrule
    Clamped-Pinned &{\parbox{5.0cm}{
    $\mathrm{cosh}\left(\mu \right)\,\sin\left(\mu \right)\\
     -\cos\left(\mu \right)\,\mathrm{sinh}\left(\mu \right) = 0$}}
      & $\sin\left(\mu - \psi(\mu)\right) =0$
      & $(4n+1)\pi/4$
    \\ \midrule
    Free-Free &{\parbox{5.0cm}{
        $\cos \left(\mu \right) \, \mathrm{cosh}\left(\mu \right)-1
        =0$ } }       & $\cos(\mu) - f(\mu) = 0$
       &  $(2n+1)\pi/2 $
    \\ \midrule
    %% Clamped-Roller &{\parbox{5.0cm}{
    %%     $\cos\left(\mu \right)\,\mathrm{sinh}\left(\mu \right) \,\\
    %%     +\,\mathrm{cosh}\left(\mu \right)\,\sin\left(\mu \right)
    %%     =0$ }  & $\sin(\mu + \psi(\mu)) = 0$
    %% \\ \midrule
    Pinned-Free & {\parbox{5.0cm}{
        $\cos\left(\mu \right)\,\mathrm{sinh}\left(\mu \right)\\
        -\,\mathrm{cosh}\left(\mu \right)\,\sin\left(\mu \right)
        =0$} } & $\sin\left( \mu - \psi(\mu) \right)=0$
       & $(4n+1)\pi/4$
    \\ \midrule
    Pinned-Pinned &  $\sin(\mu)=0$  & $\sin(\mu)=0$
    & $n\pi$
    \\ \bottomrule
	  \end{tabular}
  \caption{Euler-Bernouli: Conventional characteristic
    equations and their scaled form for common boundary
    conditions.
    \label{tab:ScaledCharEq}
  }
\end{table}
%----------------------------------------

% \clearpage % Inserted temporarily to identify the .tex files associated
%            with given text
%
%!TEX root = ../Timoshenko.tex
\pdfoutput=1
%******************************************
\subsection{Numerically stable expressions for the Timoshenko beam}
\label{sec:TB_regularization}
%******************************************
The equation for the modes of the Timoshenko beam with
$\lambda < \alpha$ are given in
Eq.~\eqref{eq: general  modeshape lambda<alpha}.  
For simplicity, we shall set
$A_1 = 1$.  Section \ref{sec:problem} discussed how when
$A_1$ and $A_2$ are not zero,
Eq.~\eqref{eq: general  modeshape lambda<alpha}
can result in numerically unstable
expressions.  
If we first focus on the equation for the
lateral displacement of the Timoshenko beam, we obtain an
expression similar to that of the Euler-Bernoulli beam
shown in Eq.~\eqref{eq:FullExpansion}.
Therefore, the
numerical stability in the displacement modes $U^*(\zeta)$ can be
resolved using the same procedure described in Section
\ref{Sec:EB_eig_forms}

We now look at the terms involved in the cross section rotation angle
$\Phi^*$ for which we have the relationship
%-------------------------------------------------
\begin{equation}
  \cosh(\mu\zeta)+A_2\sinh(\mu\zeta) =
  e^{-\mu\zeta}
  -\frac{P(\mu)}{2}\, \left(
  e^{-\mu \,\left(1-\zeta\right)}
  - e^{-\mu \,\left(1+\zeta\right)}  \right),
\end{equation}
%-------------------------------------------------
where $P(\mu)$ is given in Table \ref{tab:TBStabalizedCases} for common boundary conditions, and in Table \ref{tab:rollerTBStabalizedCases} of \ref{sec:roller_BCs} for the roller cases. 
Using this relationship in the general expression for $\Phi^*$ yields 
%-------------------------------------------------
\begin{multline}
\label{eq:phi_stable}
\Phi^* = \left(\frac{\lambda+\mu^2}{\mu}\right)
\left( e^{-\mu\zeta}
-\frac{P(\mu)}{2}\, \left(
e^{-\mu \,\left(1-\zeta\right)}
- e^{-\mu \,\left(1+\zeta\right)}  \right)
\right)
\\
  -A_3 \left(\frac{\lambda-\omega^2}{\omega}  \right)\cos(\omega \zeta)
  -R\left(1+P(\mu)e^{-\mu}\right)
  \left(\frac{\lambda-\omega^2}{\omega}  \right)\sin(\omega\zeta).
\end{multline}
%-------------------------------------------------

Therefore, using Eqs.~\eqref{eq:FullExpansion} and
\eqref{eq:phi_stable}---along with Table
\ref{tab:TBStabalizedCases} for $P$, $A_3$, and $R$---gives
the numerically stable modes for the lateral displacement
and the cross section rotation, respectively, of the
Timoshenko beam.
%% Place table for P(\mu), A_3 here 
%%%% TABLE with EXPRESSIONS for P(\mu)
\begin{table}[h]
\begin{minipage}{\textwidth}
  \centering
  % \begin{tabular}{|p{2.9cm}|p{1.25cm}|p{8.25cm}|p{1.25cm}|p{1cm}|}
    \begin{tabular}{lccc}%{|p{2.9cm}|p{8.25cm}|p{1.25cm}|p{1cm}|}
    \toprule
    Problem Definition & % Code &
    \quad\quad\quad\quad\quad $P$ &
    $A_3$ & $R$ \\
    \toprule
    Clamped-Clamped & % 'USUS' & 
  $2  \frac{ -e^{-\mu }
  + \sin (\omega ) \frac{\omega  \left(\lambda +\mu ^2\right)}{\mu 
    \left(\lambda -\omega ^2\right)}+\cos (\omega )}
{\left( 1+e^{-2\mu }\right) -2 e^{-\mu } \cos (\omega )}$ 
    &
    $\frac{\omega \,\left(\mu ^2+\lambda \right)}{\mu \,\left(\lambda -\omega ^2\right)}$ & $-1$
    %% &
    %% $\cos\left(\mu \right)\,\mathrm{cosh}
    %% \left(\mu \right)-1 = 0$
    \\[5pt] 
    \midrule  
    Clamped-Free & % 'USVM' &
    $\frac{2
  \left(-\omega e^{-\mu } \left(\lambda -\omega
  ^2\right)+\omega \left(\lambda +\mu ^2\right) \cos (\omega
  )-\mu \left(\lambda -\omega ^2\right) \sin (\omega
  )\right)}{\left(\lambda -\omega ^2\right) \left(\omega
  \left(e^{-2 \mu }-1\right)+2 \mu e^{-\mu } \sin (\omega
  )\right)} $
    & $\frac{\omega \,\left(\mu ^2+\lambda \right)}{\mu \,\left(\lambda -\omega ^2\right)}$ & $-1$
    %% &
    %% $\cos\left(\mu \right)\,\mathrm{cosh}\left(\mu \right)+1=0$
    \\[5pt]     \midrule
    Clamped-Pinned & % 'USUM' &
    $2\frac{ \left(-\mu  e^{-\mu } \left(\lambda -\omega ^2\right)+\omega  \left(\lambda +\mu
   ^2\right) \sin (\omega )+\mu  \left(\lambda -\omega ^2\right) \cos (\omega )\right)}{\mu
    \left(e^{-2 \mu }+1\right) \left(\lambda -\omega ^2\right)-2 \mu  e^{-\mu }
    \left(\lambda -\omega ^2\right) \cos (\omega )}$
    & $\frac{\omega \,\left(\mu ^2+\lambda \right)}{\mu \,\left(\lambda -\omega ^2\right)}$ & $-1$
    %% & 
    %% $\mathrm{cosh}\left(\mu \right)\,\sin\left(\mu \right)
    %% -\,\cos\left(\mu \right)\,\mathrm{sinh}\left(\mu \right) = 0$
    \\[5pt]     \midrule
    Free-Free & % 'VMVM' & 
    $-2\frac{ \omega e^{-\mu }
  \left(\lambda -\omega ^2\right)+\mu \left(\lambda +\mu
  ^2\right) \sin (\omega )-\omega \left(\lambda -\omega
  ^2\right) \cos (\omega )}
{2 \mu e^{-\mu }
  \left(\lambda +\mu ^2\right) \sin (\omega )+\omega
  \left(e^{-2 \mu }-1\right) \left(\lambda -\omega
  ^2\right)} $ & $\frac{\omega}{\mu}$ & $-\frac{\mu^2+\lambda}{\lambda-\omega^2}$
    %% &
    %% $\cos\left(\mu \right)\,\mathrm{cosh}\left(\mu \right)-1=0$
    \\[5pt]     \midrule
  \multirow{2}{*}{Pinned-Free \footnote{Note that the published version (DOI: 10.1016/j.apacoust.2019.03.015) is missing the last term in the $\Phi^*(\zeta)$ for the pinned-free case. This is a typo that has been fixed in this arXiv version.}} & 
    \multicolumn{3}{c}
  {$U^*(\zeta)=\big(\frac{\mu \cos{(\omega)}}{\omega }\big) \frac{e^{-\mu(1-\zeta)}-e^{-\mu(1+\zeta)}}{1+e^{-2\mu}}  + \sin{\omega \zeta}$}
    \\[5pt]    % \hline
    & 
    \multicolumn{3}{c}
  % $\Phi^*(\zeta)=-\frac{\lambda-\omega^2}{\omega}\cos{(\omega \zeta)}$
  {$\Phi^*(\zeta)=\big(\frac{(\lambda+\mu^2) \cos{(\omega)}}{\omega } \big)  \frac{e^{-\mu(1-\zeta)} + e^{-\mu(1+\zeta)}}{1+e^{-2\mu}} - \frac{\lambda-\omega^2}{\omega}\cos{\omega \zeta}$}  
  \\[5pt]     \midrule
  \multirow{2}{*}{Pinned-Pinned} & 
    \multicolumn{3}{c}
  {$U^*(\zeta)=\sin{(\omega \zeta)}$}
    \\[5pt]    % \hline
    & 
    \multicolumn{3}{c}
  % $\Phi^*(\zeta)=-\frac{\lambda-\omega^2}{\omega}\cos{(\omega \zeta)}$
  {$\Phi^*(\zeta)=-\frac{\lambda-\omega^2}{\omega}\cos{(\omega \zeta)}$}
    %% &
    %% $\cos\left(\mu \right)\,\mathrm{cosh}\left(\mu \right)-1=0$
    \\[5pt]     \bottomrule
    \end{tabular}
   \caption{Timoshenko beam: Stabilized cases. }
     \label{tab:TBStabalizedCases}
\end{minipage}
\end{table}
%----------------------------------------

The characteristic equations for the Timoshenko beam can
also be made numerically stable using the approach described
in Section \ref{Sec:EB_eig_forms}. 
By construction,
the numerically stable forms of the characteristic functions
(using $\sin(\psi)$ and $\cos(\psi)$) grow at most
polynomially in $\lambda$, but even that polynomial growth can
be mitigated through judicious choice of normalization.  For
instance the conventional characteristic function for the
pinned-free case is
\begin{equation}
\lambda  \mu
\left(\lambda +\mu ^2\right) \sinh (\mu ) \cos (\omega )
+\lambda  \omega 
\left(\lambda -\omega ^2\right) \cosh (\mu ) \sin (\omega ).
\end{equation}
\noindent 
Dividing through by $\sqrt{\cosh(\mu)^2+\sinh(\mu)^2}$
yields
\begin{equation}
\lambda  \mu
\left(\lambda +\mu ^2\right) \sin (\psi ) \cos (\omega )
+\lambda  \omega 
\left(\lambda -\omega ^2\right) \cos (\psi ) \sin (\omega ),
\end{equation}
which we note grows polynomially in $\lambda$  and its
functions $\mu(\lambda)$ and $\omega(\lambda)$.  For large $\lambda$,
$\mu^2\approx \lambda$ and $\omega^2\approx \lambda$, so the
above form grows approximatley at $\lambda^{2.5}$.
The one rigid body mode is captured
through the factor $\lambda$, but for $\lambda>0$, we may further
divide through by $ \lambda \sqrt{\omega \mu}  \left(\lambda +\mu^2\right)$
to obtain
\begin{equation}
\sqrt{\mu/\omega}\, \sin (\psi ) \cos (\omega )
+\sqrt{\omega/\mu} \, 
\frac{\left(\lambda -\omega ^2\right)}{\left(\lambda +\mu ^2\right)}
 \cos (\psi ) \sin (\omega )=0
\end{equation}
\noindent which grows roughly as $\lambda^0$.

For reference,
Table \ref{tab:comparison_TB_modes} contrasts the
numerically stable characteristic equations (white cells)
with the conventional expressions (shaded cells) for the
Timoshenko beam.  Note that only the case $\lambda < \alpha$
requires the numerical stabilization.
%% , but the case $\lambda
%% > \alpha$ is also included in the table for completeness.
%----------------------------------------

%-----------------------------------
%% TABLE OF CHARACTERISTIC FUNCTIONS FOR \lambda<\alpha
\begin{table}[htbp]
\centering
\begin{tabular}{p{3.28in}p{2.78in}}
\toprule
\textbf{Numerically stable expressions} & \cellcolor{light-gray} \textbf{Conventional Expressions} \\
\toprule
\multicolumn{2}{c}{ \textbf{Clamped-Clamped}} \\
\toprule
\vspace{-0.35in}
\begin{multline*}
\sin (\psi(\mu) ) \sin (\omega )
\left(\frac{\omega \left(\lambda +\mu ^2\right)}{\mu
  \left(\lambda -\omega ^2\right)}-\frac{\mu \left(\lambda
  -\omega ^2\right)}{\omega \left(\lambda +\mu
  ^2\right)}\right) \\
  +2 \cos (\psi(\mu) ) \cos (\omega )
-2 g(\mu) = 0
\end{multline*}
\vspace{-0.25in}
  % DJS 18 Aug 2018
%% $\frac{1-e^{-2\mu}}{1+e^{-2\mu}}
%% \left(\frac{\omega \left(\lambda +\mu ^2\right)}{\mu
%%   \left(\lambda -\omega ^2\right)}-\frac{\mu \left(\lambda
%%   -\omega ^2\right)}{\omega \left(\lambda +\mu
%%   ^2\right)}\right)
%% \sin (\omega )
%% +2 \cos(\omega)-4\frac{e^{-\mu}}{1+e^{-2\mu}}$
&
\cellcolor{light-gray}
\vspace{-0.35in}
\begin{multline*}
\sinh (\mu ) \sin (\omega )
\left(\frac{\omega \left(\lambda +\mu ^2\right)}{\mu
  \left(\lambda -\omega ^2\right)}-\frac{\mu \left(\lambda
  -\omega ^2\right)}{\omega \left(\lambda +\mu
  ^2\right)}\right) \\
  +2 \cosh (\mu ) \cos (\omega )-2=0
\end{multline*}
\vspace{-0.25in}
% \hline
% \multicolumn{2}{|l|}{$\boldsymbol{\lambda > \alpha$}: 
% $\sin (\theta ) \sin (\omega )
% \left(-\frac{\omega  \left(\lambda -\theta ^2\right)}{\theta
%   \left(\lambda -\omega ^2\right)}
% -\frac{\theta  \left(\lambda -\omega ^2\right)}{\omega
%   \left(\lambda -\theta ^2\right)}\right)
% -2 \cos (\theta ) \cos (\omega )+2$} 
\\
\toprule
\multicolumn{2}{c}{\textbf{Clamped-Free}} \\
\toprule
\vspace{-0.35in}
\begin{multline*}
\left( \frac{\lambda
  +\mu ^2}{\lambda -\omega ^2}+\frac{\lambda -\omega
  ^2}{\lambda +\mu ^2} \right)\cos(\psi(\mu) ) \cos (\omega
) \\ 
+\left(\frac{\omega }{\mu }-\frac{\mu }{\omega}\right)
\sin (\psi(\mu )) \sin (\omega )-2 g(\mu) = 0
\end{multline*}
\vspace{-0.15in} &
\cellcolor{light-gray}
 \vspace{-0.35in}
\begin{multline*}
\left( \frac{\lambda +\mu ^2}{\lambda -\omega
  ^2}+\frac{\lambda -\omega ^2}{\lambda +\mu ^2}\right)\cosh
(\mu ) \cos (\omega ) \\
+\left(\frac{\omega }{\mu }-\frac{\mu
}{\omega}\right) \sinh (\mu ) \sin (\omega )-2 =0 
\end{multline*}
\vspace{-0.15in}
% \\
% \hline
% \multicolumn{2}{|l|}{$\boldsymbol{\lambda > \alpha$}:
%  $\left( \frac{\theta ^2-\lambda
% }{\lambda -\omega^2} +\frac{\lambda -\omega ^2}{\theta
%   ^2-\lambda } \right) \cos (\theta ) \cos (\omega )
% -\frac{\left(\theta ^2+\omega^2\right) \sin (\theta ) \sin
%   (\omega )}{\theta \omega }+2$}
\\
\toprule
\multicolumn{2}{c}{\textbf{Clamped-Pinned}} \\
\toprule
\vspace{-0.2in}
\begin{equation*}
\frac{\mu  \left(\lambda -\omega ^2\right)}
{\omega \left(\lambda +\mu ^2\right)}\,
  \sin (\psi(\mu) ) \cos (\omega )
+\cos (\psi(\mu) )\sin (\omega ) = 0
\end{equation*}
\vspace{-0.15in}  &
\vspace{-0.25in}
\cellcolor{light-gray}
\begin{equation*}
\frac{\mu  \left(\lambda -\omega ^2\right) \sinh (\mu ) \cos (\omega )}{\omega 
  \left(\lambda +\mu ^2\right)}+\cosh (\mu ) \sin (\omega )=0
\end{equation*}
\vspace{-0.15in}
% \\
% \hline
% \multicolumn{2}{|l|}{$\boldsymbol{\lambda > \alpha$}:
% $\left(\omega ^2-\theta ^2\right) \left(\sin (\theta ) \cos (\omega )-\frac{\omega 
%    \left(\lambda -\theta ^2\right) \cos (\theta ) \sin (\omega )}{\theta  \left(\lambda
%    -\omega ^2\right)}\right)$}
\\
\toprule
\multicolumn{2}{c}{\textbf{Free-Free}} \\
\toprule
\vspace{-0.35in}
\begin{multline*}
- \left(
\frac{\mu  \left(\lambda +\mu ^2\right)}
     { \omega   \left(\lambda -\omega ^2\right)}-
\frac{\omega  \left(\lambda -\omega ^2\right)}
     { \mu  \left(\lambda +\mu ^2\right)}
     \right)\sin (\psi(\mu )) \sin (\omega ) \\
+2\cos(\psi (\mu )) \cos (\omega )-2g(\psi) = 0
\end{multline*}
\vspace{-0.25in} &
\cellcolor{light-gray}
\vspace{-0.35in}
\begin{multline*}
- \left(
\frac{\mu  \left(\lambda +\mu ^2\right)}
     { \omega   \left(\lambda -\omega ^2\right)}-
\frac{\omega  \left(\lambda -\omega ^2\right)}
     { \mu  \left(\lambda +\mu ^2\right)}
     \right)\sinh (\mu ) \sin (\omega ) \\
+2\cosh (\mu ) \cos (\omega )-2 =0
\end{multline*}
\vspace{-0.25in}
% \\
% \hline
% \multicolumn{2}{|l|}{$\boldsymbol{\lambda > \alpha$}:
% $\sin (\theta ) \sin (\omega ) \left(-\frac{\theta  \left(\lambda -\theta ^2\right)}{2 \omega
%     \left(\lambda -\omega ^2\right)}-\frac{\omega  \left(\lambda -\omega ^2\right)}{2 \theta
%     \left(\lambda -\theta ^2\right)}\right)-\cos (\theta ) \cos (\omega )+1$}
\\
\toprule
\multicolumn{2}{c}{\textbf{Pinned-Free}} \\
\toprule
\vspace{-0.35in}
\begin{equation*}
\sqrt{\mu/\omega}\, \sin (\psi ) \cos (\omega )
+\sqrt{\omega/\mu} \, 
\frac{\left(\lambda -\omega ^2\right)}{\left(\lambda +\mu ^2\right)}
 \cos (\psi ) \sin (\omega ) = 0
 \end{equation*}
\vspace{-0.15in}
%% $\lambda  \mu
%% \left(\lambda +\mu ^2\right) \sin (\psi ) \cos (\omega )
%% +\lambda  \omega 
%% \left(\lambda -\omega ^2\right) \cos (\psi ) \sin (\omega )$ DJS 18 Aug 2018
&
\cellcolor{light-gray} 
\vspace{-0.35in}
\begin{multline*}
\lambda  \mu
\left(\lambda +\mu ^2\right) \sinh (\mu ) \cos (\omega ) \\
+\lambda  \omega 
\left(\lambda -\omega ^2\right) \cosh (\mu ) \sin (\omega )=0
\end{multline*}
\vspace{-0.25in}
% \\
% \hline
% \multicolumn{2}{|l|}{$\boldsymbol{\lambda > \alpha$}:
% $\frac{\lambda \cos{(\theta)} \sin{(\omega)} (\omega^2-\theta^2)(\lambda-\omega^2)}{\theta} 
% - \frac{\lambda \cos{(\omega)}\sin{(\theta)}(\omega^2-\theta^2)(\lambda-\theta^2)}{\omega}$
% }  
\\
\toprule
\multicolumn{2}{c}{\textbf{Pinned-Pinned}} \\
\toprule
\vspace{-0.25in}
\begin{equation*}
\sin{(\omega)}=0
\end{equation*}
\vspace{-0.2in}
&  
\cellcolor{light-gray}
\vspace{-0.25in}
\begin{equation*}
\left(-\sin{(\omega)}(\mu^2+\omega^2)^2 \right)\sinh{(\mu)}=0
\end{equation*}
\vspace{-0.2in}  % conventional here
% \\
% \hline
% \multicolumn{2}{|l|}{$\boldsymbol{\lambda > \alpha$}:
% $\left(-\sin{(\omega)} (\omega^-\theta^2)^2 \right)\sin{(\theta)}$
% }  
\\
\bottomrule
\end{tabular}
\caption{Numerically stable and conventional (shaded) characteristic equations for $\lambda<\alpha$ the Timoshenko beam.}
\label{tab:comparison_TB_modes}
\end{table}
%-----------------------------------

%-------------------------------------------------------------------------------

Expressions for the characteristic functions
for $\lambda<\alpha$, $\lambda=\alpha$ and $\lambda=\alpha$
for most common boundary conditions are shown in Table
\ref{tab:TB_stable_charEq} (see \ref{sec:roller_BCs} for the roller cases).

%% TABLE FOR ALL CHARACTERISTIC FUNCTIONS.
\begin{table}[htbp]
\centering
\begin{tabular}{p{3.03in}p{3.03in}}
\toprule
\multicolumn{2}{c}{\textbf{Clamped-Clamped}} \\
\toprule
$\boldsymbol{\lambda < \alpha}$ & $\boldsymbol{\lambda > \alpha}$ \\
\midrule
\vspace{-0.35in}
\begin{multline*}
\sin (\psi(\mu) ) \sin (\omega )
\left(\frac{\omega \left(\lambda +\mu ^2\right)}{\mu
  \left(\lambda -\omega ^2\right)}-\frac{\mu \left(\lambda
  -\omega ^2\right)}{\omega \left(\lambda +\mu
  ^2\right)}\right)+ \\ 2 \cos (\psi(\mu) ) \cos (\omega )
-2 g(\mu)
\end{multline*}
\vspace{-0.25in}
%% $\frac{1-e^{-2\mu}}{1+e^{-2\mu}}
%% \left(\frac{\omega \left(\lambda +\mu ^2\right)}{\mu
%%   \left(\lambda -\omega ^2\right)}-\frac{\mu \left(\lambda
%%   -\omega ^2\right)}{\omega \left(\lambda +\mu
%%   ^2\right)}\right)
%% \sin (\omega )
%% +2 \cos(\omega)-4\frac{e^{-\mu}}{1+e^{-2\mu}}$
&
% $\alpha \sin{\omega} - \frac{2 \alpha-2\omega^2}{\omega}+\frac{2\alpha-2\omega^2}{\omega} \cos{\omega}$ &
\vspace{-0.35in}
\begin{multline*}
\sin (\theta ) \sin (\omega )
\left(-\frac{\omega  \left(\lambda -\theta ^2\right)}{\theta
  \left(\lambda -\omega ^2\right)}
-\frac{\theta  \left(\lambda -\omega ^2\right)}{\omega
  \left(\lambda -\theta ^2\right)}\right) \\
-2 \cos (\theta ) \cos (\omega )+2
\end{multline*}
\vspace{-0.25in}
\\
\midrule
\multicolumn{2}{l}{$\boldsymbol{\lambda=\alpha}$: $\alpha \sin{\omega} - \frac{2 \alpha-2\omega^2}{\omega}+\frac{2\alpha-2\omega^2}{\omega} \cos{\omega}$ } \\
\toprule
\multicolumn{2}{c}{\textbf{Clamped-Free}} \\
\toprule
$\boldsymbol{\lambda < \alpha}$ & $\boldsymbol{\lambda > \alpha}$ \\
\midrule
\vspace{-0.35in}
\begin{multline*}
\left( \frac{\lambda
  +\mu ^2}{\lambda -\omega ^2}+\frac{\lambda -\omega
  ^2}{\lambda +\mu ^2} \right)\cos(\psi(\mu) ) \cos (\omega
) + \\ \left(\frac{\omega }{\mu }-\frac{\mu }{\omega}\right)
\sin (\psi(\mu )) \sin (\omega )-2 g(\mu) 
\end{multline*} 
\vspace{-0.25in}
&
\vspace{-0.35in}
 \begin{multline*}
 \left( \frac{\theta ^2-\lambda
}{\lambda -\omega^2} +\frac{\lambda -\omega ^2}{\theta
  ^2-\lambda } \right) \cos (\theta ) \cos (\omega )
- \\ \frac{\left(\theta ^2+\omega^2\right) \sin (\theta ) \sin
  (\omega )}{\theta \omega }+2
  \end{multline*}
  \vspace{-0.25in}
   \\
  \midrule
\multicolumn{2}{l}{$\boldsymbol{\lambda=\alpha}$:  $\left(\alpha ^2-\alpha \,\omega ^2\right)\,\sin\left(\omega \right)+2\,\alpha \,\omega +\left(\omega ^3-2\,\alpha \,\omega +\frac{2\,\alpha ^2}{\omega }\right)\,\cos\left(\omega \right)-\frac{2\,\alpha ^2}{\omega } $ } \\
\toprule
\multicolumn{2}{c}{\textbf{Clamped-Pinned}} \\
\toprule
$\boldsymbol{\lambda < \alpha}$ & $\boldsymbol{\lambda > \alpha}$ \\
\midrule
$\frac{\mu  \left(\lambda -\omega ^2\right)}
{\omega \left(\lambda +\mu ^2\right)}\,
  \sin (\psi(\mu) ) \cos (\omega )
+\cos (\psi(\mu) )\sin (\omega )$ &
$\left(\omega ^2-\theta ^2\right) \left(\sin (\theta ) \cos (\omega )-\frac{\omega 
   \left(\lambda -\theta ^2\right) \cos (\theta ) \sin (\omega )}{\theta  \left(\lambda
   -\omega ^2\right)}\right)$ \\
     \midrule
\multicolumn{2}{l}{$\boldsymbol{\lambda=\alpha}$:  $\omega ^2\,\sin\left(\omega \right) $ } \\
\toprule
\multicolumn{2}{c}{\textbf{Free-Free}} \\
\toprule
$\boldsymbol{\lambda < \alpha}$ & $\boldsymbol{\lambda > \alpha}$ \\
\midrule
\vspace{-0.35in}
\begin{multline*}
- \left(
\frac{\mu  \left(\lambda +\mu ^2\right)}
     { \omega   \left(\lambda -\omega ^2\right)}-
\frac{\omega  \left(\lambda -\omega ^2\right)}
     { \mu  \left(\lambda +\mu ^2\right)}
     \right)\sin (\psi(\mu )) \sin (\omega )
+ \\ 
2\cos(\psi (\mu )) \cos (\omega )-2g(\psi)
\end{multline*}
\vspace{-0.25in} &
\vspace{-0.35in}
\begin{multline*}
\sin (\theta ) \sin (\omega ) \left(-\frac{\theta  \left(\lambda -\theta ^2\right)}{2 \omega
    \left(\lambda -\omega ^2\right)}-\frac{\omega  \left(\lambda -\omega ^2\right)}{2 \theta
    \left(\lambda -\theta ^2\right)}\right)- \\ 
    \cos (\theta ) \cos (\omega )+1
\end{multline*}
\vspace{-0.25in} \\
      \midrule
\multicolumn{2}{l}{$\boldsymbol{\lambda=\alpha}$:  $-\frac{\alpha \,\left(\alpha -\omega ^2\right)\,\left(2\,\alpha +\omega ^3\,\sin\left(\omega \right)-2\,\alpha \,\cos\left(\omega \right)-\alpha \,\omega \,\sin\left(\omega \right)\right)}{\omega } $ } \\
\toprule
\multicolumn{2}{c}{\textbf{Pinned-Free}} \\
\toprule
$\boldsymbol{\lambda < \alpha}$ & $\boldsymbol{\lambda > \alpha}$ \\
\midrule
\vspace{-0.35in}
\begin{equation*}
\sqrt{\mu/\omega}\, \sin (\psi ) \cos (\omega )
{ +
\sqrt{\omega/\mu} \, 
\frac{\left(\lambda -\omega ^2\right)}{\left(\lambda +\mu ^2\right)}
 \cos (\psi ) \sin (\omega )}
 \end{equation*}
 \vspace{-0.15in}
%% $\lambda  \mu
%% \left(\lambda +\mu ^2\right) \sin (\psi ) \cos (\omega )
%% +\lambda  \omega 
%% \left(\lambda -\omega ^2\right) \cos (\psi ) \sin (\omega )$
&
\vspace{-0.35in}
\begin{equation*}
{\frac{(\lambda-\omega^2)}{\theta} \, \cos{(\theta)} \sin{(\omega)}
  } 
-
\frac{(\lambda-\theta^2)}{\omega} \,\cos{(\omega)}\sin{(\theta)}
  \end{equation*}
  \vspace{-0.15in}
  \\
  \midrule
\multicolumn{2}{l}{$\boldsymbol{\lambda=\alpha}$: $\left(-\omega ^2\,\left(\alpha -\omega ^2\right)\right)\,\sin\left(\omega \right) $ } \\
\toprule
\multicolumn{2}{c}{\textbf{Pinned-Pinned}} \\
\toprule
$\boldsymbol{\lambda < \alpha}$ & $\boldsymbol{\lambda > \alpha}$ \\
\hline
$\sin{(\omega)}$ &
$\left(-\sin\left(\omega \right)\,{\left(\omega ^2-{\theta}^2\right)}^2\right)\,\sin\left(\theta\right) $ \\
  \hline
\multicolumn{2}{l}{$\boldsymbol{\lambda=\alpha}$: $\sin{(\omega)}$ } \\
\hline
\end{tabular}
\caption{Numerically stable expressions for characteristic
  functions for the Timoshenko beam.}
\label{tab:TB_stable_charEq}
\end{table}

%-----------------------------------------------------------------------------------

%% \input{sections/sec-asymptotic_expressions}

%% \input{sections/sec-FRF}

%!TEX root = ../Timoshenko.tex
\pdfoutput=1
\section{Results} \label{Sec:Results}
\begin{figure}[htbp]
\centering
\begin{minipage}{0.45\textwidth}
\includegraphics[width=\textwidth, trim={0 0 0.5in 0},clip]{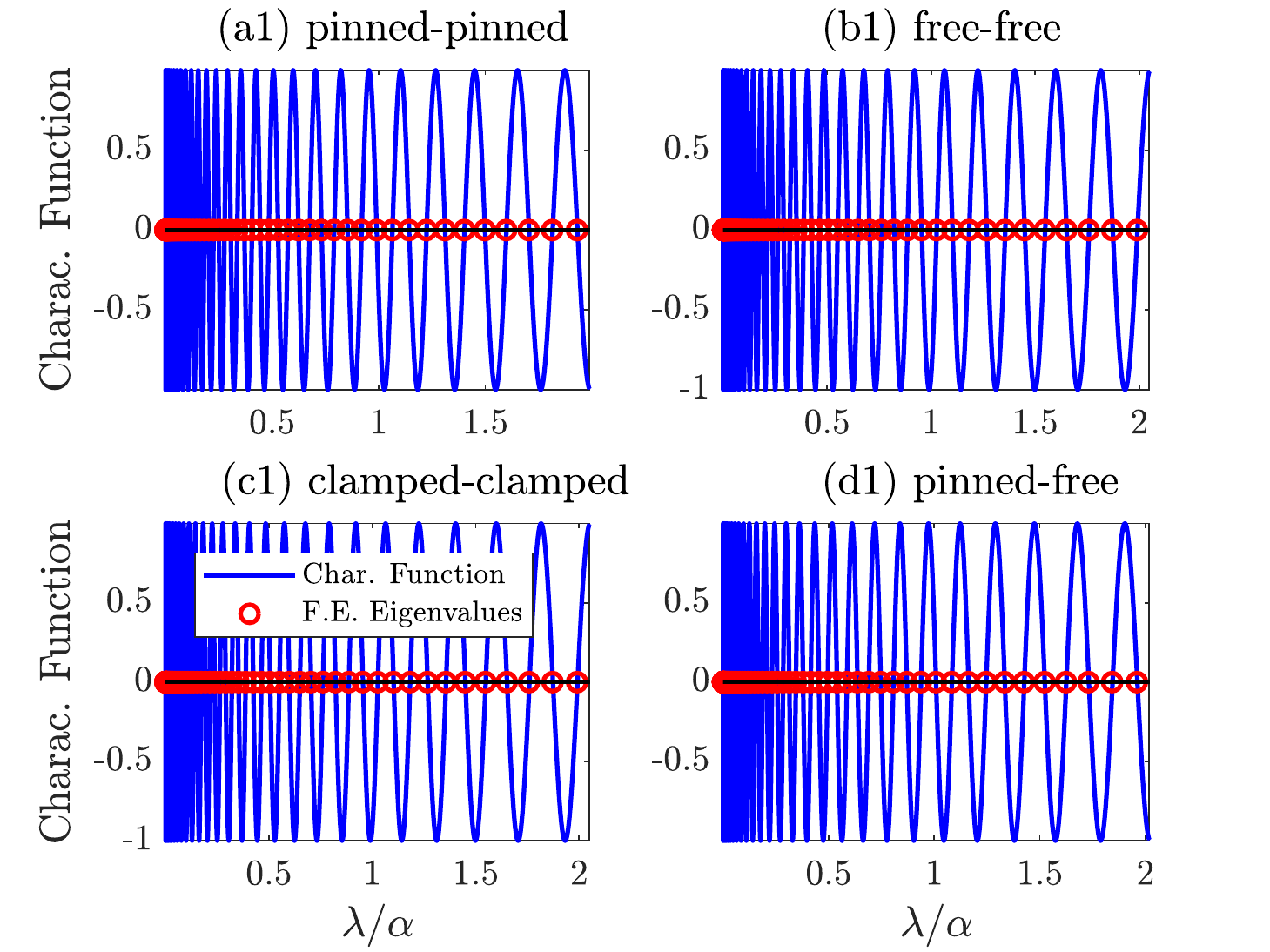}
\end{minipage}
\begin{minipage}{0.05\textwidth}
\end{minipage}
\begin{minipage}{0.45\textwidth}
\includegraphics[width=\textwidth, trim={0.5in 0 0 0},clip]{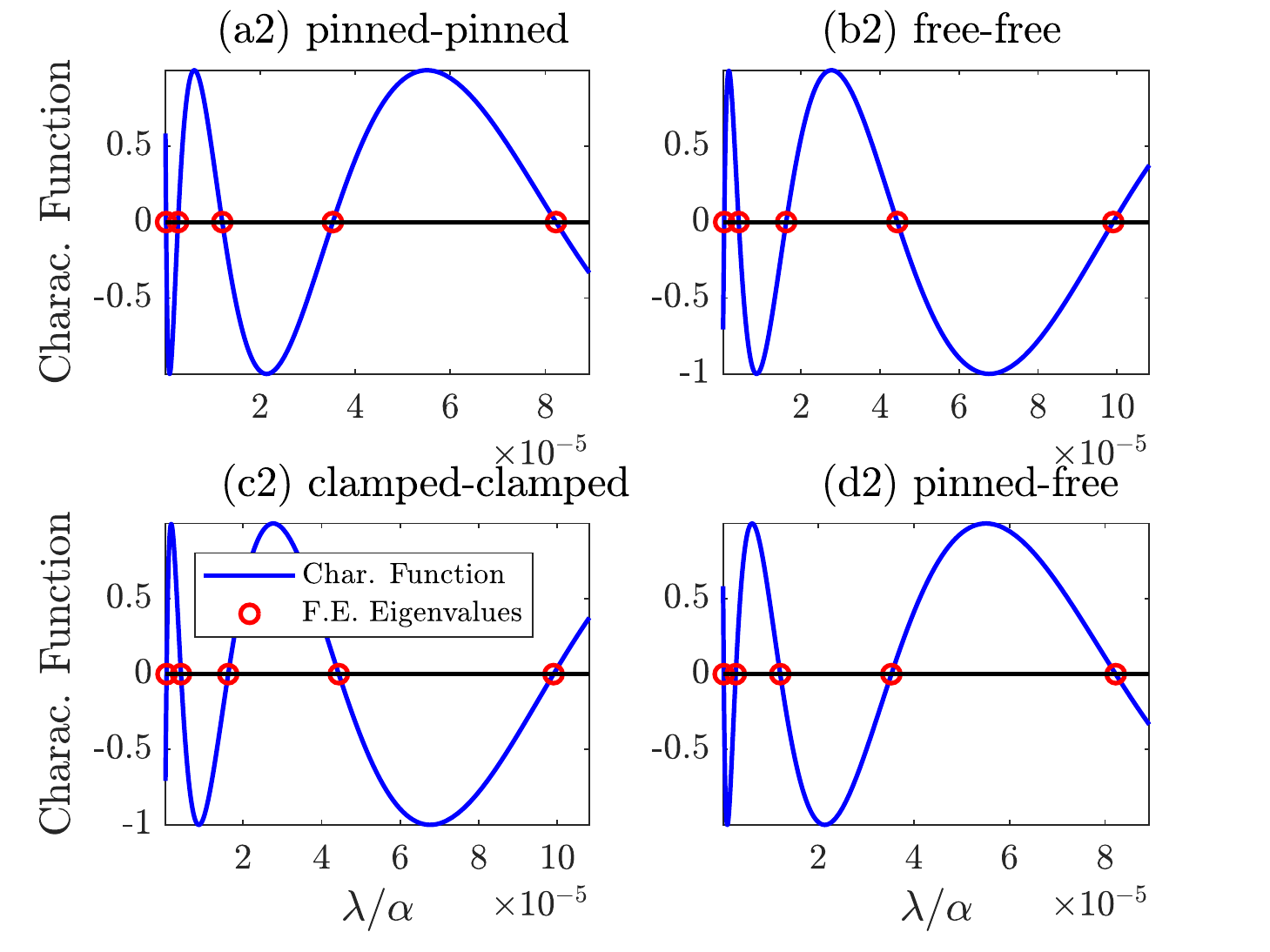}
\end{minipage}
\caption{A plot of the characteristic function for some Euler-Bernoulli beam boundary conditions obtained using the stabilized expressions (solid line) and the eigenvalues found using finite element (circles).
Plots a2--d2 are magnified versions of graphs a1--d1. }
\label{fig:EBcharEq_AnalyticVsFE}
\end{figure}

%
%--------------------------------
Plots of the scaled characteristic function of the Euler
Bernoulli beam for several combinations of boundary
condition are shown in Figure
\ref{fig:EBcharEq_AnalyticVsFE}.  As one would expect from
Table \ref{tab:StabalizedCases}, scaled characteristic
functions oscillate roughly between -1 and 1. The situation
is a bit different for the case of the characteristic
functions for the Timoshenko beam.  For those cases, the
scaled characteristic functions in general grow polynomially
in $\lambda$ for $\lambda<\alpha$ (the clamped-free case,
for instance). Further division by judiciously chosen
functions of $\lambda$ can bring the the growth of the
characteristic function down to something on the order of
$\lambda^0$ (the case of the pinned-free Timoshenko beam.)

%--------------------------------
%--------------------------------
\begin{figure}[htbp]
\centering
\begin{minipage}{0.45\textwidth}
\includegraphics[width=\textwidth, trim={0 0 0.5in 0},clip]{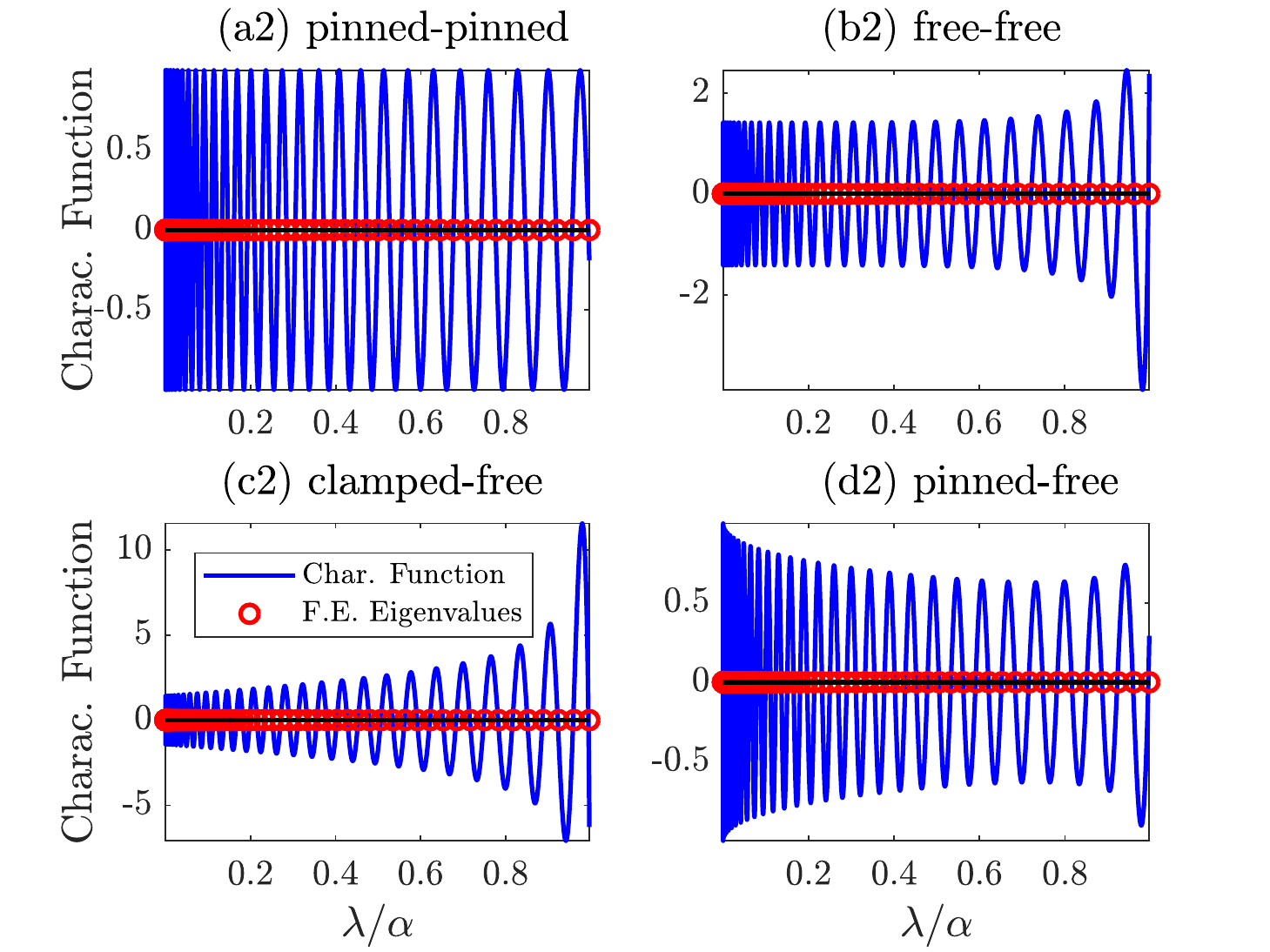}
\end{minipage}
\begin{minipage}{0.05\textwidth}
\end{minipage}
\begin{minipage}{0.45\textwidth}
\includegraphics[width=\textwidth, trim={0.5in 0 0 0},clip]{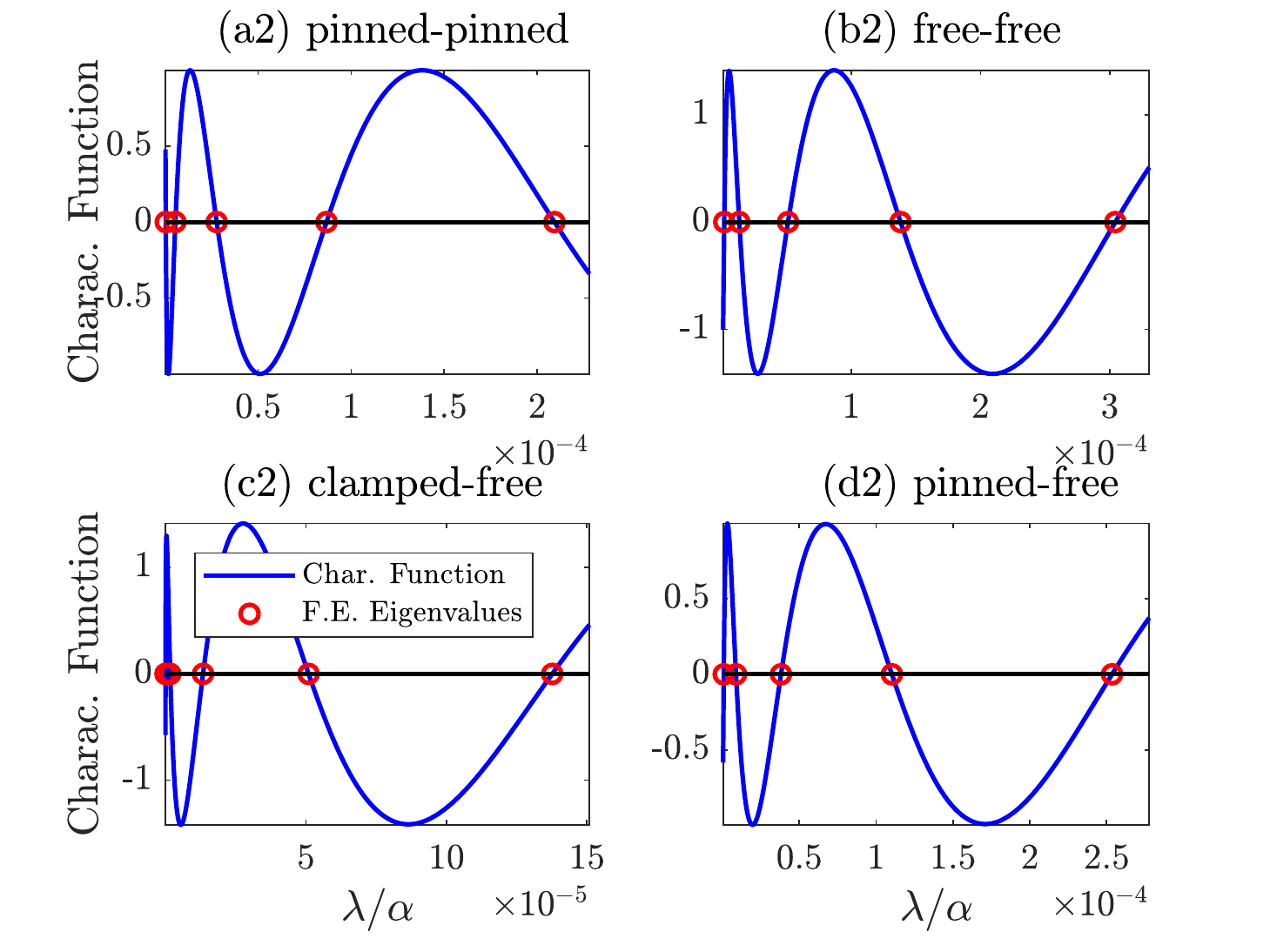}
\end{minipage}
\caption{A plot of the characteristic function for some Timoshenko beam boundary conditions obtained using the stabilized expressions (solid line) and the eigenvalues found using finite element (circles).
Plots a2--d2 are magnified versions of graphs a1--d1. }
\label{fig:TBcharEq_AnalyticVsFE}
\end{figure}

Numerical regularization makes all of these calculations as
well as the evaluation of the modes for each case
numerically well conditioned.  The next section discusses a
verification program where finite element analysis is used
to provide confidence in the correctness of the derivations
presented in this monograph and a component of this is
illustrated by the red dots in the above figures; these are
the eigenfrequencies obtained by finite element analysis and
they do coincide very closely with the zero-crossings of the
characteristic functions.

%--------------------------------

% give the results in tables;
% table 1: the eigensolutions with characteristic equations in standard form with hyperbolics
% table 2: regularized forms
% table 3: EB - standard form for 8 BCS
% table 4: EB - regularized forms
% say something about how you ideally get the same answer if you take G -> \infty, and ignoring rotary inertia
% give an exmaple for the free-free and show how it converges in the limit to the EB case
%!TEX root = ../Timoshenko.tex
\pdfoutput=1
%******************************************
% \clearpage
\section{Verification}
\label{sec:Verification-Outline}
%******************************************
The derivations in the above sections lead to exact
expressions for the characteristic functions and for the
normal modes of Euler Bernoulli and Timoshenko beams.
Though it would be good to compare these new, numerically
stable forms with the original hyperbolic expressions, such
comparison becomes impossible due to the numerical
instability of those original expressions
and to the impracticality of evaluating them.
Instead, we must
compare our analytic expressions to finite element results
for modes and frequencies.  These comparisons are not
to assess accuracy (these expressions are exact),
but to provide confidence that the derivations were
performed correctly. 
Incidentaly, though there exist asymptotic expressions for eigenfrequences of Euler-Bernoulli beams at high frequencies, they do not apply to Timoshenko beams. 
This can be seen by observing that the eigenfrequences of the two beam types can be expected to match only for cases of long ,thin beams at low frequencies to which the EB asymptotes do not apply.

Finite element formulations were used for the Euler-Bernoulli
beam and the Timoshenko beam testing the analytic
predictions for every one of the boundary condition
combinations discussed above.  The finite element results
and the analytic expressions were compared for relative
error in resonance frequency and for mutual orthonormality
with respect to the finite element mass matrix of the
finite element nodal degrees of freedom and the analytic solutions
evaluated at the nodal locations.  In all cases the error
was $0.1\%$ or less---discrepancies that can be
ascribed to mesh discretization error.  The very small
divergence between the numerical simulations and the new
analytical expressions argue that the derivations were
performed accurately.

Details of this verification process are discussed in \ref{sec:Verification}.

%
%!TEX root = ../Timoshenko.tex
\pdfoutput=1
\section{Conclusions} \label{Sec:conclusions}
The numerical difficulties associated with
expressions for natural modes for Euler
Bernoulli beams and with Timoshenko beams for high wave
number are well known and well documented in the
literature.  
Whereas Gon\c{c}alves et al
\cite{Goncalves2018} show how to overcome these problems for
the case of Euler Bernoulli beams, here we generalize the approach of \cite{Goncalves2018}, and obtain 
exact but numerically tractable expressions for the eigenvalues and the eignmodes of both the Euler-Bernoulli and the Timoshenko
beam models.
While we tabulate the resulting numerically stable expressions for four standard boundary conditions (clamped, pinned, free, and roller), we acknowledge that are there are other important boundary conditions such as spring-supported beams.
Although our approach still applies to these boundary conditions, the consideration of other boundary conditions remains a topic for future work.

Another contribution of this work is the presentation of the equations for the normal modes and the characteristic equations for beams with different boundary conditions all in the same place---and most importantly---all employing the same nomenclature.
This aims to solve a recurring problem for practitioners where it is difficult to find all that information in one place. 
Similarly, it is generally very difficult for the practitioner to find the Euler Bernoulli beam modal expressions and characteristic equations and the expressions for the corresponding Timoshenko beam using commensurate nomenclature. 
This work tackles that problem as well by presenting consistent terminology, non-dimensionalizing process, and stabilization procedures for the expressions for both beam models.

%
% \clearpage
\bibliography{TimoshenkoLit}
% \clearpage
\appendix
%!TEX root = ../Timoshenko.tex
\pdfoutput=1
%*************************************
\section{Some Detail on Timoshenko Beam Mode Derivation}
\label{sec:Apdx-Derive-Modes}
%*************************************
The forms of the eigenfunctions for lateral vibration of
Timoshenko are provided in Eqs.~\eqref{eq: general modeshape
  lambda<alpha}, \eqref{eq: general modeshape lambda=alpha},
and \eqref{eq: general modeshape lambda>alpha}.  These forms
have been presented in numerous previous papers on the
topic, but the reader might appreciate more discussion on
the derivation of these forms than is generally available in
articles, so some discussion on such derivation is presented
below.  We consider the cases of real eigenvalues ($m=\mu$)
and of complex eigenvalues ($m=\omega$) but ignore the case
of zero eigenvalue which corresonds to rigid body modes.  We begin by recalling that we are
looking for components of the mode that are represented by
$\begin{bmatrix}
U^*\\
\Phi^*
\end{bmatrix}
= \begin{bmatrix}
w_1 \,e^{m \xi}\\
w_2 \, e^{m \xi}
\end{bmatrix}$
% $U^* = w_1 e^{m \xi}$ and $\Phi^* = w_2 e^{m \xi}$
where $m$ is a solution to \eqref{eq:Delta}, the
characteristic equation of the matrix in Equation
Eq.~\eqref{eq: eigenvalue eigenfn eqn}.  Because $m$ is a
solution to that characteristic equation, the rows of the
matrix in Eq.~\eqref{eq: eigenvalue eigenfn eqn} are
proportional to each other and either row can be used to
find the ratio of $w_1$ to $w_2$; we use the first and
obtain
%-------------------------------------
\begin{equation} \label{eq:ModeRatio1}
  w_2/w_1 =  (\lambda+m^2)/m
\end{equation}
%-------------------------------------
so our modal component is
$
%% \begin{bmatrix}
%% U^*\\
%% \Phi^*
%% \end{bmatrix} =
G_1 e^{m\xi} 
\begin{bmatrix}
1 \\
(\lambda+m^2)/m
\end{bmatrix}$

Because Eq.~\eqref{eq:Delta} is even in $m$, $-m$ is also a
solution and substituting $-m$ for $m$ in the above yields a
second admissible modal component: $G_2 e^{-m\xi}
\begin{bmatrix}
	1 \\
	-(\lambda+m^2)/m
\end{bmatrix}$
%-------------------------------------
%\begin{equation}\label{eq:ModeRatio2}
%  w_2/w_1 =  -(\lambda+m^2)/m
%\end{equation}
%-------------------------------------

Combining the contributions of both of these, we express
%-------------------------------------
\begin{equation}
  \begin{bmatrix}
    U^*\\
    \Phi^*
  \end{bmatrix}
  = G_1 e^{m \xi}\begin{bmatrix}
    1 \\
    (\lambda+m^2)/m
  \end{bmatrix}
  +
   G_2 e^{-m \xi}\begin{bmatrix}
    1 \\
    -(\lambda+m^2)/m
  \end{bmatrix}
\end{equation}
%-------------------------------------

After some manipulation, this is hammered into the form
%-------------------------------------
\begin{equation}
  \begin{bmatrix}
    U^*\\
    \Phi^*
  \end{bmatrix}
  = (G_1+G_2) \begin{bmatrix}
    \frac{1}{2}\left(e^{m \xi}+e^{-m \xi}  \right) \\
    \frac{(\lambda+m^2)}{m}\frac{1}{2}\left(e^{m \xi}-e^{-m \xi}  \right) 
  \end{bmatrix}
  +
   (G_1-G_2) \begin{bmatrix}
    \frac{1}{2}\left(e^{m \xi}-e^{-m \xi}  \right) \\
   \frac{(\lambda+m^2)}{m} \frac{1}{2}\left(e^{m \xi}+e^{-m \xi}  \right)
  \end{bmatrix}
\end{equation}
%-------------------------------------

In the case of imaginary eigenvalues ($m=i \omega$), this becomes
%-------------------------------------
\begin{equation}
  \begin{bmatrix}
    U^*\\
    \Phi^*
  \end{bmatrix}
  = (G_1+G_2) \begin{bmatrix}
    \cos(\omega\xi)  \\
    \frac{(\lambda-\omega^2)}{\omega}\sin(\omega\xi)
  \end{bmatrix}
  +
   i (G_1-G_2) \begin{bmatrix}
    \sin(\omega\xi)\\
   -\frac{(\lambda-\omega^2)}{\omega} \cos(\omega\xi)
  \end{bmatrix}
\end{equation}
%-------------------------------------
We replace $ i (G_1-G_2)$ with $C$ and $(G_1+G_2)$
with $D$ to recover the corresponding trigonometric
terms in Equations
\ref{eq: general
  modeshape lambda<alpha}, \ref{eq: general modeshape
  lambda=alpha}, and \ref{eq: general modeshape
  lambda>alpha}.

In the case of real eigenvalues ($m=\mu$), this becomes
%-------------------------------------
\begin{equation}
  \begin{bmatrix}
    U^*\\
    \Phi^*
  \end{bmatrix}
  = (G_1+G_2) \begin{bmatrix}
    \cosh(\mu\xi)  \\
    \frac{(\lambda+\mu^2)}{\mu}\sinh(\mu\xi)
  \end{bmatrix}
  +
   (G_1-G_2) \begin{bmatrix}
    \sinh(\mu\xi)\\
   \frac{(\lambda+\mu^2)}{m} \cosh(\mu\xi)
  \end{bmatrix}
\end{equation}
%-------------------------------------
We replace $ (G_1-G_2)$ with $A$ and $(G_1+G_2)$ with $B$ to
recover the corresponding hyperbolic terms in Eq.~\eqref{eq:
  general modeshape lambda<alpha}.

%!TEX root = ../Timoshenko.tex
\pdfoutput=1
%*********************************
\clearpage % this is inserted for the development stage. Can be deleted
%          % later
\section{Timoshenko beam: Pinned-pinned case} \label{Sec:TB_pinned_pinned}
%*********************************
In this section, we focus on the pinned-pinned case, particularly the
possible mode shapes for the case $\lambda=\alpha$. With reference
to Eq.~\eqref{eq: general modeshape lambda=alpha}, 
the conditions that $U^*(0)=0$, $M(0)=0$, $U^*(1)=0$, and
$M(1)=0$ are captured in the matrix $B_1$ of the coefficients
$A_1, A_2, A_3, \mbox{and}, A_4$ given by 
%% Specifically, for this case and $\lambda=\alpha$, the matrix associated with the imposed boundary conditions $H_1$ and the same matrix after forward elimination $H_2$ are given by
%-------------------------------
\begin{equation}
\label{eq:pinned_pinned_elim}
B_1 =
	\begin{bmatrix}
		0 & 1        & 0                              & 1 \\
		0 & \alpha   & 0                              & \alpha - \omega^2 \\
		0 & 1        & \sin{\omega}                   & \cos{\omega} \\
		0 & \alpha   & (\alpha-\omega^2) \sin{\omega}  & (\alpha-\omega^2) \cos{\omega}
	\end{bmatrix},
	\quad
\end{equation}
After a forward elimination with pivoting, we have
\begin{equation} \label{eq:pinned_pinned_elim2}
B_2 =
	\begin{bmatrix}
		0 & 1   & 0             & 1 \\
		0 & 0   & 0             & -\omega^2 \\
		0 & 0   & \sin{\omega}  & 0 \\
		0 & 0   & 0             & 0
	\end{bmatrix}.
\end{equation}
%-------------------------------
We see no dependence on $\alpha$ in
Eq.~\eqref{eq:pinned_pinned_elim2}, and that
$\operatorname{det}B_2=0$ regardless of the value of
$\omega$.  Particularly interesting is that the first column
and the last row of $B_2$ are all zeros.  If we consider
the submatrix $\tilde{B}_2$ that excludes that row and that
column we obtain
%-------------------------------
\begin{equation}
\tilde{B}_2 =
	\begin{bmatrix}
		1   & 0             & 1 \\
		0   & 0             & -\omega^2 \\
		0   & \sin{\omega}  & 0
	\end{bmatrix};
\end{equation}
%-------------------------------
$\tilde{B}_2$ corresponds to the parameters $A_2$, $A_3$,
and $A_4$ in Eq.~\ref{eq: general modeshape lambda=alpha}
% Eq.~\eqref{eq: general modeshape lambda<alpha}
and its determinant is $\omega^2 \sin{\omega}$.  Back
substitution yields
%-------------------------------
\begin{equation}
	\begin{bmatrix}
		A_2 \\ A_3 \\A_4
	\end{bmatrix}=
	\begin{bmatrix}
		\sin{\omega}/\omega^2 \\
		1 \\
		-\sin{\omega}/\omega^2
	\end{bmatrix}
	\end{equation}
Since the characteristic equation is $\sin{\omega}=0$, we have: 
\begin{equation}
	\begin{bmatrix}
		A_2 \\ A_3 \\A_4
	\end{bmatrix}=
	\begin{bmatrix}
		0 \\  1 \\ 0
	\end{bmatrix}.
\end{equation} 
%-------------------------------
We can see from the above that the $A_1$ term, associated
with constant $\phi$, is not necessary to satisfy the
boundary conditions.  The question is now whether the
correct solution will admit an arbitrary component of
constant $\phi$.  To answer this question we consider the
strain energy of a Timoshenko beam
%-------------------------------
\begin{equation}
E = \int\limits_{0}^{L}{\left[ EI (\phi')^2 + \kappa G A (u' - \phi)^2) \right] dx}.
\end{equation}
%-------------------------------
Let  $(u_{\rm c}, \phi_{\rm c})$ be a field that 
satisfies the governing equations and the boundary
conditions, let $E_c$ be the corresponding strain energy, 
and now consider the field $(u_{\rm c}, \phi_{\rm c} +
A_1)$, where $A_1$ is a constant.  When we evaluate the
corresponding strain energy we obtain
%-------------------------------
%% \begin{equation}
%% E_{\rm c} = \frac{1}{2} \int\limits_{0}^{L}{\left[ EI
%%     (\phi_{\rm c}')^2 + \kappa G A (u_{\rm c}' - \phi_{\rm
%%       c})^2 \right] dx}.
%% \end{equation}
%-------------------------------
\begin{equation}
\bar{E} = \frac{1}{2} \int\limits_{0}^{L}{\left[ EI
    (\phi_{\rm c}')^2 + \kappa G A (u_{\rm c}' - \phi_{\rm
      c} - A_1)^2) \right] dx} = E_{\rm c} - A_1 \kappa G A
\int\limits_0^L{(u'_{\rm c} - \phi_{\rm c}) dx} +
\frac{\alpha^2 \kappa G A L}{2}.
\end{equation}
%-------------------------------
For the case of the pinned-pinned beam, the integrand above
is zero, leading to
%-------------------------------
\begin{equation}
\bar{E} = E_c + \frac{\alpha^2 \kappa GAL}{2}.
\end{equation}.
%-------------------------------
The corresponding kinetic energy is 
%-------------------------------
\begin{equation}
T = \frac{\omega_n^2}{2} \int\limits_0^L{\left[ \rho A u^2 +
    \rho I \phi^2 \right] dx} = \frac{\omega_n^2}{2}
\int\limits_0^L{\left[ \rho A u_c^2 + \rho I \phi_c^2
    \right] dx} + \frac{\omega_n^2 A_1^2 \rho I L}{2} =
\omega_n^2 \left[ \frac{M}{2} + \frac{A_1^2 \rho I L} {2}
  \right].
\end{equation}
%-------------------------------
The resulting Rayleigh quotient and its derivative with respect to $A_1$ are given by
\begin{equation}
	\label{eq:rayleigh}
	R = \omega_n^2 
	= \frac{E_{\rm c} + \frac{A_1^2 \kappa GAL}{2}}{\frac{M}{2} + \frac{A_1^2 \rho I L} {2}}, \quad \frac{\partial R}{\partial A_1} 
	= 2\left[ \frac{\kappa G A L (M + A_1^2 \rho I L) - \rho I L (2 E_{\rm c} + A_1^2 \kappa G A L)}{(M + A_1^2 \rho I L)^2}\right] A_1 
	= C(A_1) A_1,
\end{equation}
%-------------------------------
where $C(A_1) > 0$ is the function multiplying $A_1$ in
Eq.~\eqref{eq:rayleigh}.  But since $R$ must be stationary
with respect to all parameters at resonance, we conclude
that $A_1=0$.

%% There are anomalies in the results for this case associated
%% with the matrix generated by boundary conditions.  The
%% problem is pivoting on a term $\sin(\omega)$ while the
%% charactersitic equation shows that to be zero at resonance.
 \vspace*{0.25cm} 
To summarize the discussion on eigenmodes for the pinned-pinned case:
\begin{itemize}
\item for the case of $\lambda<\alpha$, the solution to the
  characteristic equation is $\sin(\omega)=0$ so the first term
  in the $A$ array is zero and the only displacement
  components left are $U^* = \sin(\omega \zeta)$,
  $\Phi^*=-\left( (\lambda-\omega^2)/\omega \right)
  \cos(\omega \zeta)$
\item for the case of $\lambda=\alpha$,
    the derivation above shows that the only
  admissible solution is the same as for $\lambda<\alpha$:
$U^* = \sin(\omega \zeta)$,
  $\Phi^*=-\left( (\lambda-\omega^2)/\omega \right)
  \cos(\omega \zeta)$
  %% I believe that the
  %% boundary conditions are inconsistent with the terms for
  %% coeffients $A$ and $B$ in Equation 13, but we have not
  %% figured out what is the characteristic equation for
  %% $\omega$ for this cases.
\item For the case of $\lambda>\alpha$
  the array $A$ would
  be written better as \\
  $A = \left\{ \begin{array}{c}
    \sin(\omega) \\ 0 \\ \sin(\theta) \\ 0
  \end{array} \right\}$ \\
  The characteristic equation asserts that either
  $\sin(\omega)=0$, or $\sin(\theta)=0$ (or both equal
  zero, which gives a null solution.) Let's examine the two cases
  \begin{enumerate}
  \item case $\sin(\omega)=0$ then the displacements are
    $U^* = \sin(\theta \zeta)$,
    $\Phi^*=-\left( (\lambda-\theta^2)/\theta \right)
    \cos(\theta \zeta)$
  \item case $\sin(\theta)=0$ then the displacements are
    $U^* = \sin(\omega \zeta)$,
    $\Phi^*=-\left( (\lambda-\omega^2)/\omega \right)
    \cos(\omega \zeta)$
  \end{enumerate}
\end{itemize}

%!TEX root = ../Timoshenko.tex
\pdfoutput=1
%******************************************
\section{Roller boundary conditions}
\label{sec:roller_BCs}
%******************************************
%----------------------------------------
\begin{table}[h]
\centering
  \begin{tabular}{|p{2.9cm}|p{3.5cm}|p{2.80cm}|p{2.0cm}|}
    \hline
    %% \multirow{2}{*}{\parbox{3.5cm}{Problem Definition \\
    %%   Code}}
    Problem Definition
    & Convent. Char. Eq. & Scaled Char. Eq & Asymp.
    \\   \hline
	Pinned-Roller & $\cos(\mu)=0 $ &  $\cos(\mu)=0 $
    & $(2n+1)\pi/2$
    \\ \hline
    Clamped-Roller & {\parbox{5.0cm}{
        $\cos(\mu)\sinh(\mu) \\ + \cosh(\mu)\sin(\mu) =  0$} }
      & $\sin\left( \mu + \psi(\mu)\right)=0$
      & $(4n-1)\pi/4$
    \\ \hline
	Roller-Free &  {\parbox{5.0cm}{
        $\cos\left(\mu \right)\,\mathrm{sinh}\left(\mu \right) \\
        +\,\mathrm{cosh}\left(\mu \right)\,\sin\left(\mu \right)$}}
        & $\sin\left( \mu + \psi(\mu)\right)=0$
    &  $(4n-1)\pi/4$
    \\ \hline
    Roller-Roller &  $\sin(\mu)=0$  & $\sin(\mu)=0$
    & $n\pi$
    \\ \hline
  \end{tabular}
  \caption{Euler-Bernoulli: Conventional characteristic equations and their numerically stabilized (scaled) form for the roller boundary conditions.
    \label{tab:rollerScaledCharEq}
  }
\end{table}
%----------------------------------------
%----------------------------------------
\begin{table}[h]
	\centering
    \begin{tabular}{|c|c|c|c|}
    \hline
    Problem Definition & 
    $P$ &
    $A_3$ & $R$ \\
    \hline
    Clamped-Roller & 
    $\frac{2\,{\mathrm{e}}^{-\mu }
      -2\,\sqrt{2}\,\sin\left(\mu +\frac{\pi }{4}\right)}
    {-{\mathrm{e}}^{-2\,\mu }
      +2\,{\mathrm{e}}^{-\mu } \,
      \sin\left(\mu \right)+1}$
    &    $-1$ & $-1$
    %% & 
    %% $\cos\left(\mu \right)\,\mathrm{sinh}\left(\mu \right) \,
    %% +\,\mathrm{cosh}\left(\mu \right)\,\sin\left(\mu \right)=0$
    \\     \hline
        Pinned-Roller  & \multicolumn{3}{c|}
        {$u=\sin(\mu \xi)$}
    %% & $\cos(\mu)=0$
    \\    \hline
    Roller-Roller &  \multicolumn{3}{c|}
    {$u=\cos(\mu \xi)$}
    %% & $\sin(\mu)=0$
    \\     \hline
    Roller-Free & 
    \multicolumn{3}{c|}
    {$u=\cos(\mu)\left( \frac{\,{\mathrm{e}}^{-\mu(1-\xi) }\,
        +\,{\mathrm{e}}^{-\mu (1+\xi) }}
        {\,{\mathrm{e}}^{-2\,\mu }+1}\right)
        + \cos(\mu \xi)$}
    %% &
    %% $\cos\left(\mu \right)\,\mathrm{sinh}\left(\mu \right)
    %% +\,\mathrm{cosh}\left(\mu \right)\,\sin\left(\mu \right)=0$
    \\     \hline
    \end{tabular}
   \caption{Euler-Bernoulli: stabilized roller cases.
   Note that only the clamped-roller case requires stabilization.}
     \label{tab:rollerStabalizedCases}
\end{table}
%----------------------------------------
%-------------------------------------------------------------------------------
\begin{table}[htbp]
\centering
\begin{tabular}{|p{3in}|p{2.3in}|}
\hline
\multicolumn{2}{|c|}{\textbf{Pinned-Roller}} \\
\hline
$\boldsymbol{\lambda < \alpha}$ & $\boldsymbol{\lambda > \alpha}$ \\
\hline
 $\cos{(\omega)} \cosh{(\mu)} $  &
$\cos{(\omega)} \cos{(\theta)}$ \\
\hline
\multicolumn{2}{|l|}{$\boldsymbol{\lambda=\alpha}$:  $\omega ^3\,\cos\left(\omega \right) $ } \\
\hline
\multicolumn{2}{|c|}{\textbf{Clamped-Roller}} \\
\hline
$\boldsymbol{\lambda < \alpha}$ & $\boldsymbol{\lambda > \alpha}$ \\
\hline
 $\sin (\omega) \cosh (\mu ) - \frac{\omega  \left(\lambda +\mu ^2\right) }{\mu  \left(\lambda-\omega ^2  \right)} \cos (\omega )\sinh (\mu ) $  &
 $\sin (\omega ) \cos{(\theta)} - \frac{\omega  \left(\lambda -\theta ^2\right) }{\theta  \left(\lambda-\omega  ^2 \right)} \cos(\omega ) \sin (\theta ) $ \\
  \hline
\multicolumn{2}{|l|}{$\boldsymbol{\lambda=\alpha}$:   $\sin\left(\omega \right)\,\left(\alpha -\omega ^2\right)-\alpha \,\omega \,\cos\left(\omega \right) $ } \\
\hline
\multicolumn{2}{|c|}{\textbf{Roller-Free}} \\
\hline
$\boldsymbol{\lambda < \alpha}$ & $\boldsymbol{\lambda > \alpha}$ \\
\hline
  $\sin\left(\omega \right)\,\cosh{(\mu)}- \frac{\omega(\lambda - \omega^2)}{\mu(\lambda+\mu^2)} \cos{(\omega)} \sinh{(\mu)} $ &
  $\sin{(\omega)} \cos{(\theta)} - \frac{\omega(\lambda-\omega^2)}{\theta(\lambda-\theta^2)} \cos{(\omega)}\sin{(\theta)} $\\
     \hline
\multicolumn{2}{|l|}{$\boldsymbol{\lambda=\alpha}$:  $\alpha ^2\,\sin\left(\omega \right)+\left(\alpha \,\omega ^3-\alpha ^2\,\omega \right)\,\cos\left(\omega \right) $  } \\
\hline
\multicolumn{2}{|c|}{\textbf{Roller-Roller}} \\
\hline
$\boldsymbol{\lambda < \alpha}$ & $\boldsymbol{\lambda > \alpha}$ \\
\hline
 $\sin{(\omega)} \sinh{(\mu)}$ &
  $\sin{(\omega)} \sin{(\theta)} $\\
      \hline
\multicolumn{2}{|l|}{$\boldsymbol{\lambda=\alpha}$:  $\left(\alpha \,\omega ^2\right)\,\sin\left(\omega \right) $ } \\
\hline
\end{tabular}
\caption{Timoshenko beam: Conventional expressions for the roller boundary condition.}
\label{tab:TB_roller_charEq}
\end{table}
%-----------------------------------------------------------------------------------

%----------------------------------------
\begin{table}[h]
  \centering
  % \begin{tabular}{|c|c|c|c|}
    \begin{tabular}{p{2.9cm}p{8.6cm}p{1.25cm}p{1cm}}
    \toprule
    Problem Definition & \qquad\qquad\qquad
     $P$ &
    $A_3$ & $R$ \\
    \toprule
    Clamped-Roller & 
    $\frac{\left(2\,\omega \,\mu ^2+2\,\lambda \,\omega \right)\,\cos\left(\omega \right)+\left(2\,\mu \,\omega ^2-2\,\lambda \,\mu \right)\,\sin\left(\omega \right)-2\,\lambda \,\omega \,{\mathrm{e}}^{-\mu }-2\,\mu ^2\,\omega \,{\mathrm{e}}^{-\mu }}{-\lambda \,\omega -\mu ^2\,\omega +\left(2\,\lambda \,\mu \,{\mathrm{e}}^{-\mu }-2\,\mu \,\omega ^2\,{\mathrm{e}}^{-\mu }\right)\,\sin\left(\omega \right)+\lambda \,\omega \,{\mathrm{e}}^{-2\,\mu }+\mu ^2\,\omega \,{\mathrm{e}}^{-2\,\mu }} $
    &    $\frac{\omega \,\left(\mu ^2+\lambda \right)}{\mu \,\left(\lambda -\omega ^2\right)}$ & $-1$
    %% & 
    %% $\cos\left(\mu \right)\,\mathrm{sinh}\left(\mu \right) \,
    %% +\,\mathrm{cosh}\left(\mu \right)\,\sin\left(\mu \right)=0$
    \\[5pt]     \midrule
\multirow{2}{*}{Pinned-Roller} & 
    \multicolumn{3}{c}
  {$U^*(\zeta)=\sin{(\omega \zeta)}$}
    \\[5pt]    % \hline
    & 
    \multicolumn{3}{c}
  % $\Phi^*(\zeta)=-\frac{\lambda-\omega^2}{\omega}\cos{(\omega \zeta)}$
  {$\Phi^*(\zeta)=-\frac{\lambda-\omega^2}{\omega}\cos{(\omega \zeta)}$}
    %% &
    %% $\cos\left(\mu \right)\,\mathrm{cosh}\left(\mu \right)-1=0$
    \\[5pt]     \midrule
\multirow{2}{*}{Roller-Free} & 
    \multicolumn{3}{c}
  {$U^*(\zeta)= \big(\frac{\cos{(\omega)} (\omega^2-\lambda)}{ (\mu^2+\lambda)}\big) \frac{e^{-\mu(1-\zeta)}+e^{-\mu(1+\zeta)}}{1+e^{-2\mu}} + \cos{(\omega \zeta)}$}
    \\[5pt]    % \hline
    & 
    \multicolumn{3}{c}
  % $\Phi^*(\zeta)=-\frac{\lambda-\omega^2}{\omega}\cos{(\omega \zeta)}$
  {$\Phi^*(\zeta)= \big(\frac{\cos{(\omega)} (\omega^2-\lambda)}{\mu }\big) \frac{e^{-\mu(1-\zeta)}-e^{-\mu(1+\zeta)}}{1+e^{-2\mu}} + \frac{\lambda - \omega^2}{\omega} \sin{(\omega \zeta)}$}
        \\[5pt]     \midrule
\multirow{2}{*}{Roller-Roller} & 
    \multicolumn{3}{c}
  {$U^*(\zeta)=\cos{(\omega \zeta)}$}
    \\[5pt]    % \hline
    & 
    \multicolumn{3}{c}
  % $\Phi^*(\zeta)=-\frac{\lambda-\omega^2}{\omega}\cos{(\omega \zeta)}$
  {$\Phi^*(\zeta)=\frac{\lambda - \omega^2}{\omega} \sin{(\omega \zeta)}$}
    \\[5pt]     \bottomrule
    \end{tabular}
   \caption{Timoshenko beam: Stabilized cases with roller boundary conditions. 
   Note that only the clamped-roller case requires stabilization. }
     \label{tab:rollerTBStabalizedCases}
\end{table}
%----------------------------------------
%
%
%
%-----------------------------------
%% TABLE OF CHARACTERISTIC FUNCTIONS, roller cases FOR \lambda<\alpha
\begin{table}[htbp]
\centering
\begin{tabular}{|p{3in}|p{3in}|}
\hline
\textbf{Numerically stable expressions} & \cellcolor{light-gray} \textbf{Conventional Expressions} \\
\hline
\multicolumn{2}{|c|}{ \textbf{Pinned-Roller}} \\
\hline
\vspace{-0.2in}
\begin{equation*}
\cos{(\omega)} \cos{(\psi(\mu))}
\end{equation*}
\vspace{-0.15in}
  % DJS 18 Aug 2018
%% $\frac{1-e^{-2\mu}}{1+e^{-2\mu}}
%% \left(\frac{\omega \left(\lambda +\mu ^2\right)}{\mu
%%   \left(\lambda -\omega ^2\right)}-\frac{\mu \left(\lambda
%%   -\omega ^2\right)}{\omega \left(\lambda +\mu
%%   ^2\right)}\right)
%% \sin (\omega )
%% +2 \cos(\omega)-4\frac{e^{-\mu}}{1+e^{-2\mu}}$
&
\cellcolor{light-gray}
\vspace{-0.2in}
\begin{equation*}
\cos{(\omega)} \cosh{(\mu)}
\end{equation*}
\vspace{-0.15in}
% \hline
% \multicolumn{2}{|l|}{$\boldsymbol{\lambda > \alpha$}: 
% $\sin (\theta ) \sin (\omega )
% \left(-\frac{\omega  \left(\lambda -\theta ^2\right)}{\theta
%   \left(\lambda -\omega ^2\right)}
% -\frac{\theta  \left(\lambda -\omega ^2\right)}{\omega
%   \left(\lambda -\theta ^2\right)}\right)
% -2 \cos (\theta ) \cos (\omega )+2$} 
\\
\hline
\multicolumn{2}{|c|}{\textbf{Clamped-Roller}} \\
\hline
\vspace{-0.15in}
\begin{equation*}
\sin (\omega) \cos{(\psi(\mu))} - \frac{\omega  \left(\lambda +\mu ^2\right) }{\mu  \left(\lambda-\omega ^2  \right)} \cos (\omega )\sin{(\psi(\mu))} 
\end{equation*}
 \vspace{-0.1in}
 &
\cellcolor{light-gray}
 \vspace{-0.3in}
\begin{equation*}
\sin (\omega) \cosh (\mu ) - \frac{\omega  \left(\lambda +\mu ^2\right) }{\mu  \left(\lambda-\omega ^2  \right)} \cos (\omega )\sinh (\mu ) 
\end{equation*}
\vspace{-0.15in}
% \\
% \hline
% \multicolumn{2}{|l|}{$\boldsymbol{\lambda > \alpha$}:
%  $\left( \frac{\theta ^2-\lambda
% }{\lambda -\omega^2} +\frac{\lambda -\omega ^2}{\theta
%   ^2-\lambda } \right) \cos (\theta ) \cos (\omega )
% -\frac{\left(\theta ^2+\omega^2\right) \sin (\theta ) \sin
%   (\omega )}{\theta \omega }+2$}
\\
\hline
\multicolumn{2}{|c|}{\textbf{Roller-Free}} \\
\hline
\vspace{-0.15in}
\begin{equation*}
\sin\left(\omega \right)\,\cos{(\psi(\mu))}- \frac{\omega(\lambda - \omega^2)}{\mu(\lambda+\mu^2)} \cos{(\omega)} \sin{(\psi(\mu))}
\end{equation*}
\vspace{-0.15in}  &
\vspace{-0.25in}
\cellcolor{light-gray}
\begin{equation*}
\sin\left(\omega \right)\,\cosh{(\mu)}- \frac{\omega(\lambda - \omega^2)}{\mu(\lambda+\mu^2)} \cos{(\omega)} \sinh{(\mu)}
\end{equation*}
\vspace{-0.15in}
% \\
% \hline
% \multicolumn{2}{|l|}{$\boldsymbol{\lambda > \alpha$}:
% $\left(\omega ^2-\theta ^2\right) \left(\sin (\theta ) \cos (\omega )-\frac{\omega 
%    \left(\lambda -\theta ^2\right) \cos (\theta ) \sin (\omega )}{\theta  \left(\lambda
%    -\omega ^2\right)}\right)$}
\\
\hline
\multicolumn{2}{|c|}{\textbf{Roller-Roller}} \\
\hline
\vspace{-0.2in}
\begin{equation*}
\sin{(\omega)} \sin{(\psi(\mu))}
\end{equation*}
\vspace{-0.15in} &
\cellcolor{light-gray}
\vspace{-0.2in}
\begin{equation*}
\sin{(\omega)} \sinh{(\mu)}
\end{equation*}
\vspace{-0.15in}
% \\
% \hline
% \multicolumn{2}{|l|}{$\boldsymbol{\lambda > \alpha$}:
% $\sin (\theta ) \sin (\omega ) \left(-\frac{\theta  \left(\lambda -\theta ^2\right)}{2 \omega
%     \left(\lambda -\omega ^2\right)}-\frac{\omega  \left(\lambda -\omega ^2\right)}{2 \theta
%     \left(\lambda -\theta ^2\right)}\right)-\cos (\theta ) \cos (\omega )+1$}
\\
\hline
\end{tabular}
\caption{Numerically stable and conventional (shaded) characteristic functions for the Timoshenko beam with roller boundary conditions and $\lambda<\alpha$. The definition for $\psi(\mu)$ can be found in Eq.~\eqref{eq:psi_def}.}
\label{tab:comparison_Roller_TB_modes}
\end{table}
%-----------------------------------
%!TEX root = ../Timoshenko.tex
\pdfoutput=1
%******************************************
\clearpage
\section{Details of the Verification Exercise}
\label{sec:Verification}
%******************************************
The derivations in the above sections lead to exact
expressions for the characteristic functions and for the
normal modes of Euler Bernoulli and Timoshenko beams.
Though it would be good to compare these new, numerically
stable forms with the the original hyperbolic expressions, such
comparison becomes impossible due to the numerical
instability of those original expressions
and to the impracticality of evalutating them.
Instead, we must
compare our analytic expressions to finite element results
for modes and frequencies.  These comparisons are not
to assess accuracy (these expressions are exact),
but to provide confidence that the derivations were
performed correctly.

\subsection{The Test Case}
We consider a beam that at low frequencies is approximated
well by an Euler Bernoulli beam but that manifests
significant shear deformation at higher frequencies, even in
the $\lambda<\alpha$ regime.

We consider a steel  beam of length 0.5 meters having a
square cross-section dimension 1.0 cm x 1.0 cm.  The
material parameters are
$\rho= 8050 \,kg/m^3$,
$E = 200 \,\mbox{x}\, 10^9 \, Pa$, and
$G= 75 \,\mbox{x}\, 10^9 \,Pa$.
The Timoshenko shear coefficient $\kappa=5/6$.
\subsection{Nondimensionalizing the Finite Element  Solutions}
One thousand elements were employed in all the finite
element results reported below.  Though this may seem
excessive, the banded nature of the structural and stiffness
matrices facilitated rapid calculation of each case.  The
finite element results were nondimensionalized an a manner
to facilitate commensurate comparison with the analytic
solutions derived in the previous section:
\begin{equation}
  \lambda^N_k = T^2 \, \lambda^{FE}_k \quad \quad
  U^N_k = U^{FE}_k/L \quad \quad
  \mbox{and} \quad \quad
  \Phi^N_k = \Phi^{FE}_k
\end{equation}
\noindent where $\lambda^{FE}_k$ is the $k^{\mbox{th}}$
eigenvalue obtained from finite element analysis, $U^{FE}_k$
is the displacement vector associated with the $k^{\mbox{th}}$
eigenmode obtained from finite element calculation, and
$\Phi^{FE}_k$ is the vector of rotations associated with the is
$k^{\mbox{th}}$ finite element eigenmode.

The $k^{\mbox{th}}$
dimensionless eigen vector is now constructed as
\begin{equation}
  (V^N_k)_{2 i-1} =  (U^N_k)_i \quad \quad
  \mbox{and} \quad \quad
  (V^N_k)_{2 i} =  (\Phi^N_k)_i 
\end{equation}
\noindent where $i \in (1,N_{\mbox{nodes}})$

The finite element mass matrix will be used to define an
inner product between eigen mode arrays, but first it must
be scaled to be consistent with the dimensionless eigenmode
arrays defined above.  Let $\Lambda$ be a diagonal matrix
whose elements are
$\left[L, \, 1, \, L, \, 1, \, \hdots \,
  L, \, 1 \right]$ and define
\begin{equation}
M_N = \Lambda \, M_{FE} \, \Lambda
\end{equation}
\noindent where $M_{FE}$ is the finite element mass matrix.
\subsection{Orthogonality}
With this new mass matrix, we define an inner product
\begin{equation}
\left(V_a, V_b\right) = V_a^T M_N V_b
\end{equation}
where $V_a$ and $V_b$ are any real column vectors of length
$2 N_{\mbox{nodes}}$.  The next step is to scale our finite
element modes so that the norm of each mode is 1.  By
construction, the eigenmodes are mutually orthogonal, and with the normalization just mentioned we now have an orthonormal
set of basis vectors:
\begin{equation}
  (V^N_m, V^N_n) = \delta_{mn}
\end{equation}
\noindent for any modes $m$ and $n$.

Next we create the corresponding vectors from the analytic
expressions for the eigenmodes.  
\begin{equation}
  (V^A_k)_{2 i-1} =  U^*(\lambda^A_k, \zeta_i) \quad \quad
  \mbox{and} \quad \quad
  (V^A_k)_{2 i}   =  \Phi^*(\lambda^A_k, \zeta_i)
\end{equation}
\noindent where $\lambda^A_k$ is the  $k^{\mbox{th}}$
root of the characteristic equation,
$\zeta_i = x_i/L$, and $\left\{ x_i \right\}$ are
the locations of the finite element nodes.  These are also
normalized so that $  (V^A_m, V^A_m) = 1$ for each mode $m$.

If the analytic expressions for the modes jibe with the
corresponding finite element modes
\begin{equation}
 (V^N_m, V^A_n) \approx \delta_{mn}
\end{equation}
\noindent for any modes $m$ and $n$.  Three natural measures
of departure of the analytic and the finite element
solutions are now
\begin{equation}
  \epsilon_{\mbox{\tiny diag}}
  =\max_{m} \left| (V^N_m, V^A_m) - 1  \right|,
  \quad \quad 
  \epsilon_{\mbox{\tiny off diag}}
  =\max_{m\ne n} \left| (V^N_m, V^A_n)  \right| ,
  \quad \quad \mbox{and} \quad \quad
  \epsilon_\lambda
  = \max_{m} \left|(\lambda^A_m-\lambda^N_m)/\lambda^A_m\right|
\end{equation}
\subsection{Finite Element Formulation}
For Euler Bernoulli beams, the standard beam formulation
(see \cite{Cook1989Finite-Element-Method}) was used, but
with the element mass matrices diagonalized using the
special lumping technique of Hinton
et. al.\cite{Hinton_etal1976Mass-Lumping-Finite-Element}.
For Timoshenko beams the formulation
of \cite{Friedman_Kosmatka19932Node-Timoshenko-Beam-Element}
was employed.

\subsection{Analytic Solution and Finite Element Comparison}

The comparison of the numerically stable analytic
eigen solutions
for the Euler Bernoulli beam with those of the corresponding
finite element analyses can be seen in Table \ref{Tab:EBBeam}.
(By analytic eigen values, we mean numerical solutions
to the characteristic equation.)

\begin{table}[htbp]
\begin{center}
\begin{tabular}{|l|c|c|c|c|} 
\hline
BC Type
& Modes
& $\epsilon_\lambda$
& $\epsilon_{\mbox{\tiny diag}}$
& $ \epsilon_{\mbox{\tiny off diag}}$ \\
\hline 
Clamped-Clamped & 65 & 0.0011 & 1.2e-08 & 8.9e-05 \\ 
\hline 
Clamped-Free & 66 & 0.0027 & 1.1e-05 & 0.0026 \\ 
\hline 
Clamped-Pinned & 65 & 0.0011 & 1.2e-08 & 8.7e-05 \\ 
\hline 
Free-Free & 65 & 0.0014 & 1e-05 & 0.0026 \\ 
\hline 
Pinned-Free & 65 & 0.0012 & 1e-05 & 0.0025 \\ 
\hline 
Pinned-Pinned & 65 & 0.0011 & 1.8e-11 & 5e-06 \\ 
\hline 
Pinned-Roller & 66 & 0.012 & 3e-07 & 0.00072 \\ 
\hline 
Clamped-Roller & 65 & 0.0011 & 1.2e-08 & 8.6e-05 \\ 
\hline 
Roller-Free & 65 & 0.0012 & 9.9e-06 & 0.0025 \\ 
\hline 
Roller-Roller & 65 & 0.0011 & 2.5e-10 & 1.4e-06 \\ 
\hline 
\end{tabular}
\caption{Comparison of numerically stable analytic solutions
with finite element results: Euler Bernoulli beam.
In each case we tested the Euler-Bernoulli beam up to roughly the same dimensionless frequency that corresponds to the Timoshenko beam.
This explains the variation in the number of considered modes (either $65$ or $66$). 
In each case, for the Euler-Bernoulli beam there were more eigenvalues below the cut-off frequency obtained from the condition described above for the Timoshenko beam, i.e., we used the same criteria ($\lambda < \alpha$) even though it does not have the same physical significance in the two beam models. 
\label{Tab:EBBeam} }
\end{center}
\end{table}

The comparison of the numerically stable analytic solutions
for the Timoshenko  beam with those of the corresponding
finite element analyses can be seen in Table
\ref{Tab:TBeam}.  The very small divergence between
the numerical simulations and the
new analytical expressions
argue that the derivations were performed
accurately.
\begin{table}[htbp]
\begin{center}
\begin{tabular}{|l|c|c|c|c|} 
\hline
BC Type
& Modes
& $\epsilon_\lambda$
& $\epsilon_{\mbox{\tiny diag}}$
& $ \epsilon_{\mbox{\tiny off diag}}$ \\
\hline 
Clamped-Clamped & 63 & 0.0023 & 8.4e-05 & 0.00019 \\ 
\hline 
Clamped-Free & 64 & 0.0022 & 0.00096 & 0.0022 \\ 
\hline 
Clamped-Pinned & 63 & 0.0023 & 6.1e-05 & 0.00023 \\ 
\hline 
Free-Free & 63 & 0.0022 & 0.0008 & 0.0019 \\ 
\hline 
Pinned-Free & 63 & 0.0022 & 0.00074 & 0.0024 \\ 
\hline 
Pinned-Pinned & 63 & 0.0024 & 4.5e-09 & 3.8e-08 \\ 
\hline 
Pinned-Roller & 63 & 0.0087 & 6e-09 & 0.00011 \\ 
\hline 
Clamped-Roller & 63 & 0.0023 & 1.4e-06 & 0.00016 \\ 
\hline 
Roller-Free & 63 & 0.0022 & 2e-05 & 0.0016 \\ 
\hline 
Roller-Roller & 63 & 0.0024 & 4.5e-09 & 1.2e-07 \\ 
\hline 
\end{tabular}
\end{center}
\caption{Comparison of numerically stable analytic solutions
with finite element results: Timoshenko beam.
Note that here we consider all the modes for which $\lambda < \alpha$ in the comparison. 
Therefore, we go up to the $64$th mode for the clamped-free boundary condition, and the $63$rd mode for all the other cases.
\label{Tab:TBeam} }
\end{table}

\end{document}